# Hipparchus’ Star Catalogues

Francesca Schironi, University of Michigan
October 24 2024

According to Pliny,[1] Hipparchus, having found a new star, decided to compile a star catalogue, with positions and magnitudes, so that it was possible to control if stars perished or were born. He even says that Hipparchus devised an instrument to check stellar magnitudes and positions. While we should not take this account at face value,[2] Ptolemy often refers to star observations by Hipparchus and gives their coordinates. In particular, discussing the precession of α Leo, Ptolemy says that the star has moved 2° and 2/3 (= 2° and 40′) along the ecliptic in the ca. 265 years from the observation of Hipparchus to when he himself observed the star at the beginning of the reign of Antoninus Pius (i.e. 137 CE), concluding that this means a precession of 1 degree per 100 years.[3] This statement implies that Hipparchus’ star observations must be dated to 129 BCE (or – 128).

In addition, the *Exegesis to the Phaenomena of Aratus and Eudoxus* (that is, what is generally known as Hipparchus’ *Commentary on Aratus)* too suggests that Hipparchus had a substantial list of stellar positions. Hence scholars have concluded that Hipparchus did indeed compose a catalogue of stars, different from the Catalogue of Simultaneous Risings and Settings that is found in the second section of the *Exegesis* (2.4.1-3.4.11). This self-standing catalogue is mostly lost; yet most recently some interesting discoveries and re-publications of previous material have brought fragments of it to the attention of scholars.

---

All translations are mine.

[1] NH 2.95: *Idem Hipparchus … novam stellam et aliam in aevo suo genitam deprehendit eiusque motu, qua fulsit, ad dubitationem est adductus, anne hoc saepius fieret moverenturque et eae, quas putamus adfixas, ideoque ausus rem etiam deo inprobam, adnumerare posteris stellas ac sidera ad nomen expungere organis excogitatis, per quae singularum loca atque magnitudines signaret, ut facile discerni posset ex eo non modo an obirent ac nascerentur, sed an omnino aliquae transirent moverenturque, item an crescerent minuerenturque, caelo in hereditate cunctis relicto, si quisquam, qui cretionem eam caperet, inventus esset.* [Hipparchus, …discovered a new star and another one which was generate in his own time, and, by its motion on the day in which it became visible, he was led to doubt of whether this often happens and whether the stars which we consider as fixed do move, and for the same reason he dared doing something that is excessive even to a god: to list stars for future generations and to check constellations with a name having created instruments through which he could mark their individual positions and magnitudes. [he did so] in order that out of this it became easy to understand non only if [stars] died or were generated but whether some would pass [each other] or move at all, as well as whether they would become bigger or smaller. [And to see] if, once the sky was left in heritance to all of us, someone was found who could take over that inheritance].

[2] See Neugebauer 1975, 283-284.

[3] Ptol. *Synt.* 7.2, 1.2, 15.6-16.2 Heiberg: παρακεχώρηκεν ἄρα ὁ ἐπὶ τῆς καρδίας τοῦ Λέοντος εἰς τὰ ἑπόμενα τοῦ διὰ μέσων τῶν ζῳδίων μοίρας β̅ Γβ τῶν ἀπὸ τῆς τοῦ Ἱππάρχου τηρήσεως ἐτῶν μέχρι τῆς ἀρχῆς Ἀντωνίνου, καθ' ἣν μάλιστα καὶ ἡμεῖς τὰς πλείστας τῶν ἀπλανῶν παρόδους τετηρήκαμεν πέντε που καὶ ἑξήκοντα καὶ διακοσίων συναγομένων, ὡς ἐκ τούτων τὴν τῆς μιᾶς μοίρας εἰς τὰ ἑπόμενα παραχώρησιν ἐν ἑκατὸν ἔγγιστα ἔτεσιν γεγενημένην εὑρῆσθαι, καθάπερ καὶ ὁ Ἵππαρχος ὑπονενοηκὼς φαίνεται, δι' ὧν φησιν ἐν τῷ Περὶ τοῦ ἐνιαυσίου μεγέθους οὕτως· “Εἰ γὰρ παρὰ ταύτην τὴν αἰτίαν αἵ τε τροπαὶ καὶ ἰσημερίαι μετέβαινον εἰς τὰ προηγούμενα τῶν ζῳδίων ἐν τῷ ἐνιαυτῷ μὴ ἔλασσον ἢ ἑκατοστὸν μιᾶς μοίρας, ἔδει ἐν τοῖς τριακοσίοις ἔτεσιν μὴ ἔλασσον ἢ γ̅ μοίρας αὐτὰ μεταβεβηκέναι” [Therefore the [star] on the heart of the Lion has moved 2° 2/3 towards the trailing direction [i.e. east] of the line in the middle of the zodiacal sings [i.e. the ecliptic] in the years of Hipparchus’ observation until the beginning of Antoninus’ [kingship], in which period I have observed the majority of the fixed [stars]’ motions, after ca. 265 years have passed, so that from these [data] we found that the backward motion toward the west consists of one degree in ca. 100 years, just like Hipparchus too seems to have suspected, in [the passage] in *On the Length of the Year* in which he speaks as follows: “For if, due to this [reason] the solstices and equinoxes were moving, in one year, toward the western direction of the zodiac signs not less than 1/100 of one degree, in 300 years they would have moved not less that 3 degrees”].

Remains of Hipparchus’ catalogue have been found in excerpts from papyri and medieval manuscripts. The most famous and most important evidence is the tiny fragment which has been recently recovered in a palimpsest (*Codex Climaci Rescriptus*, henceforth: CCR), dating to the fifth or sixth century CE, which provides some coordinates for stars in the Northern Crown.[4] Secondly, *P.Aberd.* 12, a papyrus dating to the second-third century CE, preserves another list of constellations, in particular the description of the Small Bear and the Dragon. This too has just been republished.[5] Three other fragments of this catalogue have been identified in the so-called *Aratus Latinus* (henceforth: AL), that is, the Latin scholia to Aratus.[6] In particular, there is an entry on the Small Bear, one on the Great Bear, and one on the Dragon.[7] These three fragments were briefly studied by Neugebauer,[8] but did not generate much interest at that time; however, the two recent publications of CCR and of *P.Aberd.* 12 have made them much more relevant. The aim of this article is to analyze all these fragments and compare them to the evidence provided by the *Exegesis*, to assess their value and what they can tell us about Hipparchus’ engagement with stars and star catalogues.

## 1 CCR and the Northern Crown

I will start from the fragment in CCR, since it is the best preserved and can be used as a ‘model’ to study the others. The tiny fragment contains the description of the Northern Crown:

> Ὁ Cτέφανος ἐν τῷ βορείῳ ἡμιcφαιρίῳ κείμενος κατὰ μῆκος μὲν ἐπέχει μ̊ θ̅ καὶ δ̅′ ἀπὸ τῆς α̅ μ̊ τοῦ Cκορπίου ἕως ι̅ <καὶ> δ̅′ μ̊ τοῦ αὐτοῦ ζῳδίου. Κατὰ πλάτος δ’ ἐπέχει μ̊ ς̅ C καὶ δ̅′ ἀπὸ μ̅θ̅ μ̊ ἀπὸ τοῦ βορείου πόλου ἕως μ̊ ν̅ε̅ C καὶ δ̅′.
> Προηγεῖται μὲν γὰρ ἐν αὐτῷ ὁ ἐχόμενος τοῦ λαμπροῦ ὡς πρὸς δύςιν ἐπέχων τοῦ Cκορπίου τῆς α̅ μ̊ τὸ ἥμιcυ. Ἔςχατος δὲ πρὸς ἀνατολὰς κεῖται ὁ δ′ ἐχόμενος ἐπ’ ἀνατολὰς τοῦ λαμπροῦ ἀςτέρος [. . .] τοῦ βορείου πόλου μ̊ μ̅θ̅· νοτιώτατος δὲ ὁ γ′ ἀπὸ τοῦ λαμπροῦ πρὸς ἀνατολὰς ἀριθμούμενος ὃς ἀπέχει τοῦ πόλου μ̊ ν̅ε̅ C καὶ δ̅.[9]

> ι̅ <καὶ> δ̅′: Gysembergh, Williams and Zingg, ι̅δ̅′ CCR
> μ̊: Gysembergh, Williams and Zingg, μ̣έ̣ς̣ῃ̣ς̣ CCR
> ὁ δ′ ἐχόμενος· Gysembergh, Williams and Zingg, ὁ δὲ ἐχόμενος CCR

> The Crown which lies in the northern hemisphere occupies in length 9 degrees and ¼, from **the first degree of the Scorpion** to 10 degrees <and a> ¼ (CCR: 14 and a half) of the same zodiacal sign. In breadth it occupies 6 degrees and ¾, from 49 degrees from the northern pole to 55 degrees and 3/4.
> For in it the star (β CrB) which is toward the west, next to the bright one (α CrB), leads, occupying **half of the first degree of the Scorpion**. The last one, toward the east, is the one which is fourth (ι CrB) (CCR: the one which is) to the east of the bright one (α CrB) […]
> [. . .] 49 degrees [from] the northern pole. Southernmost is the third [star] (δ CrB), counted from the bright one (α CrB) toward the east, which is 55 degrees and ¾ from the northern pole.

[4] Gysembergh, Williams and Zingg 2022.

[5] Savio 2023, 257-274. The papyrus was first published by Turner 1939, 14, who however did not identify it correctly (he labeled it as ‘astronomical computation?’); see also Luiselli 2011, 151-152, who concludes that the genres of this fragment is difficult to determine (commentary to Aratus? star catalogue? Isagogic text to Aratus?). The much improved edition by Savio interprets it as a catalogue of stars which could be used both for didactic purposes but also for time reckoning on the basis of the stars’ rising; on this, see below.

[6] On the Aratus Latinus, see Martin 1956, 42-51 and Le Bourdellès 1985; edited by Maass 1898, 172-306.

[7] Edited by Maass 1898, 183-184, 186-187, 189.

[8] Neugebauer 1975, 288-291. For this study, I could also consult the notes by Neugebauer in his notebooks now at the Institute for Advanced Studies in Princeton. I would like to thank Alex Jones for sending me scans of the relevant pages.

[9] Text from Gysembergh, Williams and Zingg 2022, 386.

I have underlined the passages which are the result of emendation by the editors and added below the critical apparatus to show what CCR reads. In the translation I have provided translations for both the emended passages and for the reading in CCR. I have also printed in bold values that are in disagreement.

As explained by the editors, this description gives the boundaries of the constellation, in two steps:

1) the E-W and N-S extensions, with the total number of degrees occupied by the constellation in each direction, followed by the precise indication of the starting and ending point of the arc.
2) stars marking those boundaries, with equatorial coordinates.

The identification of the stars marking the boundaries is relatively easy since the Northern Crown is a quite stable constellation, with 8 stars, easy to identify. In particular, in this entry many of these stars are recognized by using α CrB, the brightest star of the constellation, as the reference point. The western boundary is marked by β CrB, the eastern boundary is marked by ι CrB—the latter star is however identified as such only thanks to an emendation. Otherwise, what CCR reads (ὁ δὲ ἐχόμενοϲ ἐπ᾽ ἀνατολὰϲ τοῦ λαμπροῦ ἀϲτέροϲ = the one which is to the east of the bright one) would point to γ CrB. The southern boundary is marked by δ CrB; the star marking the northern boundary is missing, since the manuscript is damaged, but has been restored by the editors as π CrB on the basis of the coordinates provided.

The numerical values instead are a bit more problematic. First, given the structure, one would expect the starting and ending points of both directions in section one (E-W extension, N-S extension) to match the coordinates of the four stars marking the E, W, N, S boundaries. This indeed occurs for the N-S boundary. However, for the E-W boundary, the value of the eastern limit in the first part is the result of emendation by the editors; in the second part, on the other hand, the star marking that limit (the 'restored' ι CrB) is not accompanied by its coordinates, probably due to a lacuna in the text. As for the western limit, the value in the first section does not match (even if by little) the one provided for β CrB.

Clearly this entry has suffered a considerable amount of damage. However, if we accept their emendations, these are the data provided by the fragment; in square brackets I have put data that are missing in the text but can be deduced from the context:

Boundaries

- W-E extension (right ascension): 9º1/4, from **Scorpion 1º** to Scorpion 10º1/4 (where the value for the easternmost boundary is result of emendation and the westernmost boundary is different from the one given below for the star marking the westernmost boundary)
- N-S extension (polar distance): 6º3/4, from 49º to 55º3/4 from the North Pole.

Stars Marking the Boundaries

- W: the star (β CrB) which is toward the west, next to the bright one (α CrB): **Scorpion 0.5º = 210º 30′** (right ascension) (this value is different from the one given above for the westernmost boundary).
- E: the one which is fourth (ι CrB) to the east of the bright one (α CrB): [Scorpion 10º1/4 = 220º 15′ (right ascension )].[10]

[10] The RA of ι CrB is missing in the text but is easily deducible from the boundaries given above.

- N: [π CrB]:[11] 49º (polar distance) [= decl. +41º].
- S: the third [star] (δ CrB), counted from the bright one (α CrB) toward the east: 55º3/4 (polar distance) [= decl. +34º 15′].

The coordinates connected with those four stars can be compared to those calculated for 129 BCE:

- W: β CrB: right ascension 210º 02′
- E: ι CrB: right ascension 219º 13′
- N: π CrB: decl. +40º 48′
- S: δ CrB: decl. +34º 16′

This comparison shows that this catalogue with the emendations proposed by Gysemberg and al. reflects the coordinates of the time of Hipparchus.

Among these coordinates, only one can be compared to the data of the *Exegesis*: the RA of β CrB and indeed it matches the value given by Hipparchus in the latter work:

> **3.5.8** Τὸ δὲ δεύτερον ὡριαῖον διάϲτημα ἀφορίζει περὶ τὴν ἀρχὴν τοῦ Cκορπίου…τοῦ δὲ ϲτεφάνου ὁ προηγούμενοϲ τοῦ λαμπροῦ ὡϲ τριακοϲτὸν μέροϲ ὥραϲ ὑπολείπεται τοῦ διὰ τῶν πόλων κύκλου, ὁμοίωϲ δὲ καὶ ὁ μέϲοϲ τῶν ἐν τῷ μετώπῳ τοῦ Cκορπίου τριῶν λαμπρῶν ἀϲτέρων.
>
> The second one-hour interval, around the beginning of the Scorpion, is delimited by [the following stars]: … of the Crown, the [star] that precedes the bright one (β CrB) remains behind the circle through the poles by ca. 1/30 of an hour (i.e., 2 minutes)

This is the section in the *Exegesis* which lists stars marking the 24 1-hour meridians. β CrB is listed as 'almost' marking the meridian passing by the beginning of the Scorpion on the equator, that is, 210º; Hipparchus says that it is behind this meridian by 1/30h (= 2m), which results in a right ascension of 210° 30', just like CCR maintains, at least when providing the coordinates for the star. However, in the first section, when CCR gives the boundaries of the constellations, the value is different, since there the RA is given as the first degree of the Scorpion, not half of it (see however footnote 13 for a possible solution of this contradiction).

This matching of one coordinate (albeit with the caveats just discussed) could simply suggest that the two values, in CCR and in the *Exegesis*, are the result of precise measurements around the same time, but not necessarily by the same person. However, one detail that, in my view, points to the Hipparchan origin of this fragment is the way right ascension is expressed. For example, in ἀπὸ τῆϲ α̅ μ̊ τοῦ Cκορπίου, "from the first degree of the Scorpion", Hipparchus speaks of degrees within "zodiacal signs" along the celestial equator. Yet zodiacal signs are a division of the ecliptic, not of the celestial equator. As I have discussed elsewhere,[12] Hipparchus seems to be adapting the system of "ecliptic longitudes" of the Babylonians to the celestial equator, thus using an "ecliptic (or zodiacal) nomenclature" for sections of the celestial equator. While this system sounds odd to us, it was probably perfectly natural to Hipparchus, who reasoned in terms of spherical geometry: since both the ecliptic and the equator are great circles, he probably felt entitled to use the Babylonian division into 12 zodiacal signs (each 30 degrees long) and transfer it to the equator. As far as we know, this system is attested only in Hipparchus' *Exegesis*—hence the fact that we find it in this fragment from CCR strongly suggests a Hipparchan origin.

[11] This is the supplement of the authors; see Gysembergh, Williams and Zingg 2022, 392, n.10.

[12] See Schironi (forthcoming)-b. Other scholars had already noted that Hipparchus divided the equator into 12 zodiacal signs; see Delambre 1817, 1, 172 ("Il divise l'équateur en douze signes"); Neugebauer 1975, 278 ("Still stranger is the division of the equator and of its parallels in 'zodiacal signs' which are denoted by the same names as the signs in the ecliptic"); Toomer 1978, 217 ("He frequently uses an odd form of right ascension. Thus he says that a star 'occupies three degrees of Leo along its parallel circle'. This means that each small circle parallel to the equator is divided into twelve 'signs' of 30°, and thus the right ascension of the star is 123°"); Duke 2002, 428.

In addition, also the way the stars are described, with reference to the brightest star, α CrB, matches what we see in the *Exegesis*. The only case that does not correspond to how Hipparchus describes the same star in the *Exegesis* is actually the 'emended description' to have ι CrB ('fourth' from α CrB). In the *Exegesis* Hipparchus describes it as "the dimmer one, which is also the last one of the trailing", which seems a more obvious way to describe it (figure 1). Even so, Hipparchus is not always consistent in describing the same stars even in the *Exegesis* itself, so such a difference should not be taken as a proof that the entry is CCR is not Hipparchan in origin.

A synoptic table listing these four stars with Ptolemy's and Hipparchus' descriptions together with the latter's coordinates in the *Exegesis* and the data in CCR will make the situation clearer (I have underlined the points of agreement between the two sources):

| Ptolemy | Star | Hipparchus in Exegesis | Coordinates in Exegesis | CCR |
|---|---|---|---|---|
| | | | | |
| ὁ προηγούμενος πάντων<br>The [star] which precedes all | β CrB | **2.5.2** ὁ προηγούμενος τοῦ λαμπροτάτου<br>the one that precedes the brightest one (α CrB)<br><br>**2.5.9; 3.5.8** ὁ προηγούμενος τοῦ λαμπροῦ<br>the [star] that precedes the bright one (α CrB) | RA (**3.5.8**): Beginning of the Scorpion (210º), behind by 1/30h = 2m = 210º 30′ | ὁ ἐχόμενος τοῦ λαμπροῦ ὡς πρὸς δύσιν<br>the star which is toward the west, next to the bright one (α CrB)<br><br>RA: Scorpion 0.5º = 210º 30′ |
| ὁ ἔτι τούτῳ ἑπόμενος καὶ βορειότερος<br>The [star] that further follows the latter (θ CrB) and is more northern | π CrB | | | lacuna for star description;<br><br>Polar distance: 49º [= decl. + 41º] |
| ὁ ἔτι τούτῳ ἐγγὺς ἑπόμενος<br>The [star] which further follows the latter (γ CrB) by close distance | δ CrB | **3.3.6** ὁ τρίτος ἀπὸ τοῦ λαμπροῦ ἐπὶ τὰ ἑπόμενα<br>the third [star] [starting] from the bright one (α CrB) toward the trailing directions [i.e. east] | | ὁ γʹ ἀπὸ τοῦ λαμπροῦ πρὸς ἀνατολὰς ἀριθμούμενος<br>the third [star], counted from the bright one (α CrB) to the east<br><br>Polar distance: 55º 45′ [= decl. + 34º 15′] |
| ὁ πᾶσι τοῖς ἐν τῷ Στεφάνῳ ἑπόμενος<br>The [star] which trails after all the [stars] in the Crown | ι CrB | **2.6.2** ὁ ἀμαυρότερος καὶ ἔσχατος ὢν τῆς ἑπομένης περιφερείας<br>the dimmer one, which is also the last one of the trailing arc<br><br>**2.5.2** ὁ βορειότερος τῶν ὡς πρὸς ἀνατολὰς καὶ ἄρκτους κειμένων τοῦ λαμπροτάτου.<br>the more northern [star] of those which lie to the north-east of the brightest one (α CrB). | | ὁ δʹ ἐχόμενος ἐπʹ ἀνατολὰς τοῦ λαμπροῦ<br>the [star] which is fourth to the east of the bright one (α CrB)<br><br>RA: Scorpion 10º1/4 = 220º 15′ |

The one element in CCR that does not reflect the Hipparchan usage as it emerges from the *Exegesis* is the way the E-W and N-S extensions are expressed. These are arcs expressed in degrees and in CCR ordinal numbers are used, just like for the specific coordinates of the four stars marking those boundaries.[13] On the other hand, in the *Exegesis* when giving an extension for an arc (mostly the ecliptic but also once for the equator), indicating the beginning and the ending point in degrees (i.e. with ἀπὸ…ἕωϲ…), Hipparchus uses always ordinals (e.g. "the ecliptic rises from the 10th degree of Virgo to the 13th of Virgo"). In addition, in the *Exegesis*, with ordinal numbers the only fraction used is 'half of' a degree and not more complex fractions, as instead occurs here. More complex fractions are used in the *Exegesis* only for stellar coordinates, and there they are all expressed with cardinals. Moreover, even the 'half' of a degree, when expressed in ordinals in the *Exegesis*, is always μέϲη μοῖρα and never τὸ ἥμιϲυ μοίραϲ as we have here in the second phrase under discussion (α̅ μ̊ τὸ ἥμιϲυ). Thus, the way CCR expressed the E-W and N-S extension is not in line with what we find in the *Exegesis*. [14]

To conclude, while the data, including the 'zodiacal' terminology for the right ascension, seem to go back to Hipparchus, some specific details in the way coordinates are expressed are different from the practice we can securely attribute to Hipparchus since it is attested in the *Exegesis*, the only text by him which has reached us by direct tradition. CCR thus seems to have preserved a fragment of a catalogue of stars going back to Hipparchus, but we cannot exclude that the wording might have been changed in the course of the transmission. As we will see with the other fragments preserved in AL and in *P.Aberd.* 12, all these fragments seem to have undergone various types of corruption in their tradition.

## 2 The Aratus Latinus and the Great Bear

Despite the problems discussed above, the fragment in CCR is quite well preserved, at least when we compare it to the other two sources, AL and *PAberd* 12, which provide more problematic texts and less data. In particular, AL is written in a very idiosyncratic Latin, which is highly corrupted and grammatically flawed.[15] In addition, these scholia have come down in two recensions, the 'original' Aratus Latinus and the so-called *Aratus Latinus recensio interpolata* or *Revised Aratus Latinus*, none of which is actually better than the other.[16]

The three fragments in AL are very corrupted texts, both in terms of Latin and numbers; for this reason, Neugebauer only presented an outline with the content, but not a real translation.[17] Here, however, because I am actually interested in discussing the transmission of these texts (see below, § 5), I have decided to present the text of the *Aratus Latinus*, with a translation of what should have been the original text behind the garbled Latin we now face. In my attempt to make (some) sense of the text, in the Latin I have indicated with asterisks passages or words which are clearly corrupted; in square brackets I have suggested possible readings, both in terms of Latin,[18] or astronomical data, mostly concerning numbers. Dots in square brackets indicate passages in which something is obviously missing. In the translation I have followed the same criteria, but I have only indicated corrections (both with asterisks and suggested readings in square brackets) only when it comes to astronomical data.

[13] They are all cardinals because no article is preposed to those letter numerals. The only exceptions are the two articles τῆϲ in front of α̅ μ̊: Ὁ Στέφανοϲ ἐν τῷ βορείῳ ἡμιϲφαιρίῳ κείμενοϲ κατὰ μῆκοϲ μὲν ἐπέχει μ̊ θ̅ καὶ δ̅΄ ἀπὸ τῆϲ α̅ μ̊ τοῦ Σκορπίου ἕωϲ ι̅ <καὶ> δ̅΄ μ̊ τοῦ αὐτοῦ ζῳδίου. … Προηγεῖται μὲν γὰρ ἐν αὐτῷ ὁ ἐχόμενοϲ τοῦ λαμπροῦ ὡϲ πρὸϲ δύϲιν ἐπέχων τοῦ Σκορπίου τῆϲ α̅ μ̊ τὸ ἥμιϲυ. As Alex Jones, who kindly discussed the question with me remarked, the article here can be explained if we assume that in both cases the correct value was 'half of the first degree' or 0° 30': the value (30′) is indeed a cardinal but the expression of this fraction incorporates an ordinal; this might be a way of expressing exact positions less than 1° in the absence of a zero. If this is correct, we should emend the first sentence, adding τὸ ἥμιϲυ and readjusting the total length. If the extension was from 0° 30' to 10° 15' , the total extension would be 9º 3/4, which, expressed in Egyptian fractions, would be 9º 1/2 ¼; if so, we should admit that the symbol for ½ dropped out. The restored sentence would be: Ὁ Στέφανοϲ ἐν τῷ βορείῳ ἡμιϲφαιρίῳ κείμενοϲ κατὰ μῆκοϲ μὲν ἐπέχει μ̊ θ̅ καὶ <∠> δ̅΄ ἀπὸ τῆϲ α̅ μ̊ < τοῦ ἡμίϲεωϲ> τοῦ Σκορπίου. However, the problems connected with this entry make it necessary to be very cautious with any attempted emendation.

[14] On Hipparchus' use and expressions of coordinates, see Schironi forthcoming, Chapter 5.

[15] For a detailed linguistic study of this language, see Le Bourdellès 1985, 114-150, 155-160.

[16] See Le Bourdellès 1985, 71-81.

[17] Neugebauer 1975, 288-290. On the other hand, in his Notebooks Neugebauer had tried a translation similar to mine.

As will become clear by the following analysis, in theory many other emendations would be possible—especially when it comes to numbers. However, what I am trying to do here is to show that *in theory* those 'wild' texts can be shown to preserve original Hipparchan data similar to those preserved by CCR, *if* we allow for a certain amount of corruption, which can be however shown to be *paleographically possible*. The explanations of why I have proposed certain emendations in the astronomical data will follow the texts, since they imply discussing the content and comparing it with the evidence from Hipparchus' *Exegesis*.

The entry relative the Great Bear in AL is the following one:[19]

> Maior Arcturus, ad aquilonem adiacens, in longitudine quidem recipit […] aequinoctiales sortes *ciclus* [pro: circuli?] partium *trecentorum sexaginta quinque* [pro: trecentorum sexaginta?]. A Geminis […] *aequinoctium* [pro: aequinoctio[20]?] manente usque ad Vergilias sortes *V* [pro: IV?].
> in latitudine quidem distat sortes *XXIII* [pro: XXI semis?], quas sortes X et VIII semis ab aquilonio polo usque ad sortes XL.
>
> […] quod iuxta antecedit […]. super ipsam in gutture praeclara stella distans a Geminis sortes X et VIII semis.
> novissimus quidem ad orientales adiacet distans sortes *XXX* [pro: IV?] […].
> item aquilonius est qui super humeros, qui distat ab aquilonio polo sortes XVIII semis.
> australis qui in duobus retrorsis pedibus, qui *distant* [pro: distat?] a polo aquilonio sortes XL.
> restant *X et duo centum* [pro: XXI?] sortes et pars.
>
> The Great Bear, which lies toward the north, in length occupies […] equinoctial degrees of the *365* [rather: 360] parts of the [equinoctial] circle. From the Twins […], with the fixed equinox [? Aries 0º?], to [the rising of?] the Pleiades (i.e. the autumn equinox = 180º),[21] *5* [rather: 4?] degrees.
> In width it extends *23* [rather: 21 ½?] degrees, namely, 18 ½ degrees from the north pole to 40 degrees [from the north pole].
>
> […] which/because it precedes a little [...]. Above it there is the bright star in the throat, which is 18 ½ degrees distant from the Twins.
> The last [star] towards east lies *30* [rather: 4?] degrees distant […?].

---

[18] Only for clearly obvious errors which concern important terms to understand the technical content. In other cases, in which the Latin is not correct, yet clear enough or involving only easy morphological changes, I have not suggested the more correct form in order not to cloud the text too much.

[19] For the original edition (with no indications of possible lacunae or errors), see Maass 1898, 183-184; for a summary of the content, see Neugebauer 1975, 288-289.

[20] This is the reading of the *Revised Aratus Latinus*.

[21] As suggested by Neugebauer 1975, 289. In Geminus' parapegma, the Pleiades rise in the evening around the autumn equinox; see Evans and Berggren 2006, 234 In his notebooks Neugebauer suggested that *Virgiliae* might be a corruption for *Virgo*. This might be; however, in that case, the text would read "Maiden, 5º degrees (= 155º)", which is incorrect; hence I have preferred to adopt the other option, even if it might be an odd way to express a right ascension of 180º.

And the northern [star] is the one which is on the shoulders, which is 18 ½ degrees distant from the northern pole.
The southern [star] [is] the one which is in the two hind feet, which is 40 degrees distant from the northern pole.
There remain *112* [rather: 21?] degrees and a fraction.

Despite the textual problems, we can reconstruct the following data (missing data are in square brackets):

Boundaries

- W-E extension (right ascension): 360- [##] degrees; from the Twins [## degrees] to 185º (i.e. 180º + 5º) [rather: 184º (i.e. 180º + 4º)?]
- N-S extension (polar distance): 23 degrees [rather: 21.5 degrees?], from 18º 1/2 to 40º from the north pole

Stars Marking the Boundaries

- W: the bright star in the throat: Twins 18º ½ (right ascension) = RA 78º 30′.
- E: the last [star] towards east: [zodiacal sign] 30º [rather: 4º?] (right ascension) = RA ?
- N: the [star] on the shoulders: 18º 1/2 (polar distance) [= decl. + 71º 30′].
- S: the [star] in the two hind feet: 40º (polar distance) [= decl. + 50º].

To compare these data with those in the *Exegesis*, it is easier to start from the stars marking the boundaries.

In the *Exegesis* Hipparchus does not mention any star in the throat of the Great Bear. But he mentions a star in the snout: o UMa.[22] Assuming a slight change in the anatomical part, due mostly to translation issues, o UMa would work well since its right ascension in 129 BCE was 78º 37′.[23]

The star marking the eastern boundary as well as its precise right ascension are lost in the manuscript. Neugebauer suggests η UMa.[24] Indeed, in the *Exegesis* Hipparchus considers it the last star of the Bear to the east; he also gives a right ascension:

1.5.10 …ὁ γὰρ ἐν ἄκρᾳ τῇ οὐρᾷ τῆϲ Ἄρκτου, ὅϲ ἐϲτιν ἔϲχατοϲ τῶν πρὸϲ ἀνατολὰϲ τῶν ἑπτά, κατὰ παράλληλον τῷ ἰϲημερινῷ κύκλον ἐπέχει Χηλῶν μοίραϲ δ̅.

For the [star] within the tip of the tail of the Bear (η UMa), which is the last one of the seven [stars] toward the east, occupies 4 degrees of the Claws along the circle parallel to the equator.

If the identification is correct, then we must assume that the right ascension given by AL (i.e. 30 degrees from…), on top of being lacunose is also corrupted, since we expect something like "4 degrees of the Claws", that is, 184º. Indeed, the '30 degrees' (XXX sortes) in the Latin could be explained as a paleographical mistake in the original Greek, by which the original δ̅ (4) was written λ̅ (30), being the two letters both triangular-shaped letters, and so easy to confuse. At any rate, η UMa is indeed the eastern limit of the Great Bear. Its right ascension in 129 BCE was 184º 22′, thus matching Hipparchus' value.

[22] Hipp. 2.6.10 ὅ τε ἐν τῷ ῥύγχει τῆϲ Μεγάληϲ Ἄρκτου the [star] in the snout of the Great Bear (o UMa);

[23] Ptolemy lists o UMa as being on the tip of the snout (just like Hipparchus) and then he has also two stars in the neck, τ UMa and 23(h) UMa, but they do not work here; see Neugebauer 1975, 289.

[24] Neugebauer 1975, 289.

The northern limit is marked by the 'star on the shoulders'. According to Hipparchus it is α UMa,[25] which indeed in 129 BCE had a declination of 71º 28′; this value matches the AL text almost perfectly since a polar distance equal to 18º 30′ implies a declination of +71º 30′.

The 'star in the hind feet' marking the southern limit is more difficult to determine. Ptolemy mentions λ UMa and μ UMa in the left back foot and ν UMa and ξ UMa in the right back foot. The identification of the back feet of the Bear in Hipparchus is more difficult to determine. He might have had the same shape and stars in mind, but his description of those stars does not allow to be certain except for λ UMa, which he considers the "the leading [star] of those in the back feet" (3.4.10).[26] Thus, to check the data in AL, it is better to compare the declinations of these four stars in 129 BCE, which were as follows:

ν UMa = decl. +43º 41′
ξ UMa = decl. + 42º 30′
λ UMa= decl. + 51º 30′
μ UMa = decl. + 50º 16′

The most southern one of these four is ξ UMa, whose declination is however ca. 8 degrees off the one presented by AL (decl. + 50º); similarly, ν UMa is also too much to the south. What we expect, on the basis of the declination given by AL, is either λ UMa or μ UMa. If we trust the description of AL as going back to Hipparchus, we need thus to conclude that, unlike Ptolemy, Hipparchus did not include ν UMa and ξ UMa in the Great Bear—which is actually quite possible since they are very much to the south (figure 2).

Choosing between λ UMa and μ UMa is more difficult because both declinations could fit. However, μ UMa would work best. In fact, even if Hipparchus does not mention μ UMa in the *Exegesis*, he describes λ UMa as the leading [star] of those in the back feet (**3.4.10** ὁ ἡγούμενος τῶν ἐν τοῖϲ ὀπιϲθίοιϲ ποϲίν). This implies that there was a 'trailing' star—which is indeed μ UMa. On the other hand, AL seems to assume only one star in the hind feet—yet this might be due to a corruption of the text.

To conclude, by putting together the data in AL and those emerging from the *Exegesis,* we have now the following 'revised' data for the stars marking the Great Bear:[27]

Stars Marking the Boundaries: Identifications and Data for 129 BCE

- W: the bright star in the throat (ο UMa): Twins 18º ½ (right ascension) = RA 78º 30′ ≈ RA 129 BCE: 78º 37′
- E: the last [star] towards east (η UMa): [Claws] 4º (right ascension) = RA 184º = *Exegesis* ≈ RA 129 BCE: 184°22′
- N: the [star] on the shoulders (α UMa): 18º 1/2 (polar distance) [= decl. + 71º 30′] ≈ decl. 129 BCE: 71º 28′
- S: the [star] in the two hind feet (μ UMa): 40º (polar distance) [= decl. + 50º] ≈ decl. 129 BCE + 50º 16′

These identifications allow to improve also on the first section, with the overall boundaries of the Great Bear. The W-E extension has a slightly different RA for the eastern boundary: 185º rather than 184º but, as we have seen, 184º is the RA of η UMa is also attested in the *Exegesis*. It is indeed rather easy to suggest that what is now “sortes V” (5 degrees to be added to the 180º) originally was “sortes IV” in Latin; if instead we think of a corruption in the Greek original, we should suppose a change from δ to ε –not impossible but less likely perhaps. As for the N-S extension, the correction of the transmitted total of 23 degrees into 21 ½ degrees is supported by the subtraction of the two limits (i.e. from 18º 1/2 to 40º). Paleographically we would have to suppose a change from “XXI semis” into “XXIII” in Latin or (perhaps more likely) a change from κι ∠ into κγ (with iota and the symbol for ½ to be taken for a majuscule gamma, Γ). In fact, a further support that the original number for the N-S extension was 21 ½ could be found in the final note after the stars marking the N and S limit: what now reads as “there remain *112* degrees and a fraction” could be instead read as “there remain *21 degrees and a fraction”, where an original XXI was somehow turned into X et duo centum.[28]

[25] Hipp. 3.2.11 τῆϲ τε Ἄρκτου ὁ ἐπὶ τῶν ὤμων αὐτῆϲ of the Bear, the [star] on its shoulders (α UMa).

[26] See Schironi (forthcoming)-a, ad 3.2.12.

[27] The same identifications were suggested by Neugebauer 1975, 289 and Gysembergh, Williams and Zingg 2022, 388. Here I have explained why these identifications are correct on the basis of the evidence of Hipparchus' *Exegesis.*

The reconstructed version of AL thus pretty much matches Hipparchus’ data in the *Exegesis*, as the following table also shows:

| Ptolemy | Star | Hipparchus in Exegesis | Coordinates in Exegesis | AL |
|---|---|---|---|---|
| ὁ ἐπ' ἄκρου τοῦ ῥύγχους<br>The [star] on the tip of the snout | ο UMa | **2.6.10** ὅ τε ἐν τῷ ῥύγχει τῆς Μεγάλης Ἄρκτου<br>the [star] in the snout of the Great Bear | | the star in the throat<br><br>RA: Twins 18º ½ (= 78º 30′) |
| ὁ τρίτος καὶ ἐπ' ἄκρας τῆς οὐρᾶς<br>The third one and on the tip of the tail | η UMa | **1.4.4** ἡ ἄκρα οὐρά<br>The tip of the tail<br><br>**1.5.10** …ὁ … ἐν ἄκρᾳ τῇ οὐρᾷ τῆς Ἄρκτου, ὅς ἐστιν ἔσχατος τῶν πρὸς ἀνατολὰς τῶν ἑπτά<br>the [star] within the tip of the tail of the Bear, <u>which is the last one of the seven [stars] toward the east</u><br><br>**1.5.13, 3.2.13, 3.2.14** ὁ ἐν ἄκρᾳ τῇ οὐρᾷ<br>the [star] in the tip of the tail | RA: **(1.5.10):** 4 degrees of the Claws (cf. also **1.5.13**) = 184º | <u>the last [star] towards east;</u><br><br>RA: ** 30º [rather: Claws 4º = 184º] |
| τῶν ἐν τῷ τετραπλεύρῳ ὁ ἐπὶ τοῦ νώτου<br>Of the [stars] in the quadrangle, the one on the back | α UMa | **3.2.11** τῆς τε Ἄρκτου {τοῦ πλινθίου} ὁ ἐπὶ τῶν ὤμων αὐτῆς<br>of the Bear, <u>the [star] on its shoulders</u><br><br>**1.5.2** ὁ βορειότερος ἀστὴρ τῶν δύο τῶν ἡγουμένων ἐν τῷ πλινθίῳ<br>the more northern star of the two leading ones (α, β UMa) in the quadrangle [[in the head according to Eudoxus and Aratus; cf. also **1.6.3**]]<br><br>**1.5.4** ὁ βορειότερος τῶν ἡγουμένων ἐν τῷ πλινθίῳ.<br>the more northern of the leading [stars] in the quadrangle | RA (**1.5.4**): a bit less than 3 degrees of the Lion = 121º<br><br>RA (**1.5.8**): 3rd degree of the Lion = 122º<br><br>RA (**1.6.3**): 2nd degree of the Lion = 121º | <u>The [star] on the shoulders;</u><br><br>Polar distance: 18º 1/2 [= decl. + 71º 30′] |
| τῶν ἐν τῷ ὀπισθίῳ ἀριστερῷ ἀκρόποδι ὁ προηγούμενος<br>Of the [stars] in the left back foot, the leading one | λ UMa | **3.4.10** ὁ ἡγούμενος τῶν ἐν τοῖς ὀπισθίοις ποσίν<br>the leading [star] of those in the back feet | | |
| ὁ τούτῳ ἑπόμενος<br>The one which follows the latter (λ UMa) | μ UMa | | | The [star] in the two hind feet;<br><br>Polar distance: 40º [= decl. + 50º] |

[28] This is a suggestion in Neugebauer 1975, 289.

# 3 The Aratus Latinus, PAberd 12, and the Small Bear

The Small Bear is described by both AL and *PAberd* 12. The entry of AL is the following:[29]

> Minor Arcturus, qui ad aquilonem adiacet, in longitudine quidem habet quartam partem aequinoctialis circuli et sortes *septem* [pro: septemdecim?].
> a latere autem habet sortes *una* [pro: quattuor?] *semis* [pro: duas quintas partes?], quae ab octo sortibus minus sunt duodecim *pars* [pro: sortes?] et *duo quinta sors* [pro: duae quintae partes?].
>
> antecedunt autem illam quae deducunt *XII* [pro: II?], de quibus in humerum habent principatum de Sagittario.
> novissima ad orientem adiacet ad summum habens Piscem sortes XVII.
> item *aquilonium* [pro: aquilonius?] duobus, quibus in humero distat aquilonio polo sortes *XL* [pro: VIII?].
> restant aequinoctiales.
> australis quoque est, qui in cauda.
> reliqui vero duo nominantur Circenses, eo quod in circuitu perambulent.
>
> The Little Bear, which lies toward the north, in length has the fourth part of the equinoctial circle and *7* [rather: 17?] degrees.
> And from the side (i.e. in width) it has *1* [rather: 4?] degrees and *½* [rather: 2/5?], which are 12 degrees and 2/5 less of 8 degrees.
>
> There are *12* [rather 2?][30] [stars] which lead in it, of which those in the shoulder occupy the beginning of the Archer.
> The very last one to the east lies to the extreme [limit], occupying the Fish, 17 degrees.
> The northern one of the two [stars] in the shoulder is *40* [rather: 8?] degrees away from the northern pole.
> There remain […] equinoctial […].
> The southern [star] is also the one in the tail.
> The two remaining [stars] are called 'Circenses' because they go around a circle.[31]

With my proposed corrections we have the following data:

<u>Boundaries</u>

- W-E extension (right ascension): ¼ of an equinoctial circle (= 90 degree) + 7 [rather: 17] degrees = 97 [rather: 107] degrees
- N-S extension (polar distance): 1 [rather: 4?] degrees and ½ [rather: 2/5?]: from 8º to 12º and 2/5 from the north pole

<u>Stars Marking the Boundaries</u>

[29] For the original edition (with no indications of possible lacunae or errors), see Maass 1898, 186-187; for a summary of the content, see Neugebauer 1975, 289-290.

[30] According to Neugebauer 1975, 290, who quotes Toomer's suggestion, this means that there are 12 (or 10) stars outside the Small Bear; the only parallel is a scholium to Germanicus which says that there are 10 stars in front of the Small Bear; see Robert 1878, 59, l.10. This note is not present, however, in the scholium as edited by Dell'Era (Rome 1979) (III, p. 318). In fact, these 10 stars outside the Small Bear are unattested otherwise; thus, I suggest that the text might be corrupted, and the number is actually 2, indicating the two most western [i.e. leading] stars of the Small Bear, which are β UMi and γ UMi (see discussion below).

[31] Neugebauer 1975, 291 suggests that these are γ UMi and η UMi since in ca. 129 BCE they had the same polar distance (ca. 11.5º). Similar *Sch. Germ.* in Robert 1878, 59, l.13-14; in Dell'Era edition (III, p. 318) however, the phrasing is different: Praedicti autem II reliqui uocantus Choreutae de circuito.

- W: stars in the shoulder: Archer 0° (right ascension) = RA 240°
- E: the last to the east: Fishes 17° (right ascension) = RA 347°
- N: the northern one of the two [stars] in the shoulder: 40° [rather: 8°?] (polar distance) [= decl. + 50 [rather: + 82°?]
- S: the [star] in the tail […polar distance missing]

As is clear, among other problems, here too there are inconsistencies with numbers between the coordinates in the first part (general W-E and N-S limits) and the coordinates of the stars marking those limits. Again, it is better to start from the stars marking the boundaries. In the *Exegesis* Hipparchus describes the Small Bear as mostly made of a quadrangle with a tail—which is exactly what also Ptolemy does. The stars forming the quadrangle are: β γ η ζ UMi. Some of them are also defined anatomically by Hipparchus:

- β UMi: the head of the Small Bear for Eudoxus and Aratus (Hipp. 1.4.2)
- γ UMi: the more southern star on the front feet for Eudoxus and Aratus (Hipp. 1.4.2)[32]
- η UMi: never defined anatomically in the *Exegesis*
- ζ UMi: at the juncture of the body with the tail the tail of the Small Bear (Hipp. 1.11.16)

Of the other three stars listed by Ptolemy in the Small Bear (α UMi, δ UMi and ε UMi), in the *Exegesis* Hipparchus only mentions α UMi, which is the last star of the tail.[33] This could be indeed 'the star in the tail' marking the southern boundary in AL. In fact, a further confirmation of this identification comes from Ptolemy, who in *Geogr.* I, 7, 4 gives the following information regarding the most southern star of the Small Bear for Hipparchus:

> Παραδίδοται δὲ ὑπὸ τοῦ Ἱππάρχου τῆϲ μικρᾶϲ Ἄρκτου ὁ νοτιώτατοϲ, ἔϲχατοϲ δὲ τῆϲ οὐρᾶϲ ἀϲτὴρ ἀπέχειν τοῦ πόλου μοίραϲ ιβʹ καὶ δύο πέμπτα.
>
> the most southern star of the Small Bear is given by Hipparchus, as the last star of the tail being distant from the pole by 12 degrees and 2/5.

The polar distance provided by Ptolemy matches the one provided in the first part as the southern limit, so the identification is secure. In addition, it does correspond to the time of Hipparchus, since in 129 BCE α UMi had a declination equal to +77° 32′, which is only one degree off from 78° 36′, the declination corresponding to a polar distance of 12° 24′.

In fact, α UMi is also the eastern boundary of the constellation and in 129 BCE its right ascension was 347° 34.′ In the *Exegesis* Hipparchus gives its RA as the 18th degree of the Fishes, a value expressed in ordinal numbers which, when translated into cardinal numbers, corresponds to Fishes 17°,[34] exactly as reported in AL. Indeed, at the time of Hipparchus α UMi rather than being 'polaris', was the most southern (and eastern) star of the Small Bear (figure 3).

As for the other stars, AL mentions as western limit the "stars in the shoulder" and as northern limit "the northern one of the two [stars] in the shoulder". This description implies that there are two stars in the shoulders, one to the north of the other, and both to the west side of the constellation. Given the shape of the Small Bear the only possibility is that these stars were β UMi and γ UMi. As Hipparchus reports, in Eudoxus' and Aratus' asterisms, they were the head (β UMi) and the front feet (γ UMi) of the Bear; if the text preserved by AL goes back to Hipparchus, he then seems to have had a bigger picture for this animal shape since now these two stars are placed on the shoulders (see below). The identification with β UMi and γ UMi is indeed correct, since the celestial coordinates in 129 BCE for these stars match those reported in AL. In 129 BCE β UMi had a declination of 82º 04′ and a right ascension equal to 240º 51′; γ UMi has a right ascension of 240º 04′. Indeed, the beginning of the Archer (= 240º) is the right ascension provided by the AL for the stars marking the western limit. Moreover, the same right ascension is provided in the *Exegesis*, where β UMi, γ UMi are also considered to be at beginning of the Archer (Hipp. 3.5.10).[35] As for β UMi a declination of 82º 04′, corresponds to a polar distance of ca. 8º; this is the value present in the first part of the AL. When listing the star itself AL has "40 degrees from the North Pole". As already suggested by Neugebauer,[36] the 40 is explicable with an error in the Greek version, whereby the original η (8) was transcribed as μ (40). This evidence allows for a rather secure identification of the stars with their coordinates: [37]

---

[32] Hipp. 1.4.2 οἱ γὰρ λαμπρότατοι καὶ ἡγούμενοι τῶν ἐν τῷ πλινθίῳ ταύτηϲ ἀϲτέρων, ὧν ὁ μὲν βορειότερόϲ ἐϲτι κατ' αὐτοὺϲ[32] ἐπὶ τῆϲ κεφαλῆϲ, ὁ δὲ νοτιώτεροϲ ἐπὶ τῶν ἐμπροϲθίων ποδῶν [For the brightest and leading ones of the stars in its quadrangle (β UMi and γ UMi), of which the more northern (β UMi) according to them is on the head and the more southern (γ UMi) is on the front feet]

[33] Hipp 1.5.19: τοῦ ἐν ἄκρᾳ τῇ οὐρᾷ τῆϲ Ἄρκτου [in the tip of the tail of the [Small] Bear].

[34] For this conversion, see Schironi forthcoming, chapter 5.

Stars Marking the Boundaries: Identifications and Data for 129 BCE

- W: stars in the shoulder (β UMi, γ UMi): Archer 0º (right ascension) = RA 240º = *Exegesis* ≈ RA 129 BCE 240º 51′( β UMi)/ 241º 04′ (γ UMi)
- E: the last to the east (α UMi): Fishes 17º (right ascension) = RA 347º = *Exegesis* ≈ RA 129 BCE: 347º 45′
- N: the northern one of the two [stars] in the shoulder (β UMi): 40º [rather: 8º] (polar distance) [= decl. + 82º] ≈ decl. 129 BCE +82º 04′
- S: the [star] in the tail (α UMi): 12º 2/5 = 12º 24′ (polar distance) [= decl. + 78º 3/5 = +78º 36′] = Hipp. In Ptol. *Geogr.* ≈ decl. 129 BCE +77º 32′

After these identifications, it is rather easy to fix what remains to be fixed in the first part of the entry with the total E-W and N-S extensions. The W-E extension is given as a total of 97º in the first description, but such a sum does not match the two right ascensions given for the stars marking those boundaries (beginning of Archer and 17 degrees of the Fishes). The difference between the latter two right ascensions is 107º. This is why I suggested reading "in longitudine quidem habet quartam partem aequinoctialis circuli et sortes septemdecim [in length has the fourth part of the equinoctial circle and 17 degrees]. The error is easily explicable as the omission for the symbol of 10, either in Latin (X) or in the Greek original $\overline{\iota}$).

As for the N-S extension, the two limits (N = 8 degrees and S = 12 2/5 degrees away from the pole) are indeed correct; what is not correct is the total length, since the full N-S width is ca. 4º 2/5, which is what I propose reading in the text—even if it is admittedly difficult to figure out who the corruption could have generated either in Latin (from "sortes quattuor/IV duas quintas partes" to "sortes una semis" vel ) or in Greek (from "$\overline{\delta}$ μοι καὶ δύο πεμπτημόρια" to "$\overline{\alpha}$ μοι καὶ ∠" or "$\overline{\alpha}$ μοι καὶ τὸ ἥμισυ"?)—the problem being not the full number (IV can be easily corrupted into I just as $\overline{\delta}$ into $\overline{\alpha}$) but the fraction.

A synoptic table again will help to assess the combined data:

[35] At 1.7.11 he says that β UMi is 'the end of the Scorpion', which is however almost the same

[36] Neugebauer 1975, 290.

[37] The same identifications were suggested by Neugebauer 1975, 290 and Gysembergh, Williams and Zingg 2022, 388.

| Ptolemy | Star | Hipparchus in Exegesis | Coordinates in Exegesis | AL |
|---|---|---|---|---|
| ὁ ἐπ' ἄκρας τῆς οὐρᾶς<br>The [star] on the tip of the tail<br><br>Ptolemy, Geogr. I, 7, 4] Polar distance acc. to Hipparchus: 12º 24′ | α UMi | **1.2.12** ἡ οὐρά<br>the tail [of the Small Bear]<br><br>**1.5.19** ὁ ἐν ἄκρᾳ τῇ οὐρᾷ<br>the [star] the tip of the tail<br><br>**1.6.4** ὁ ἔϲχατοϲ καὶ λαμπρότατοϲ ἀϲτὴρ<br>the last and the brightest star | RA: (**1.6.4**): 18th degree of the Fishes | The [star] in the tail<br><br>RA = Fishes 17º = RA 347º<br><br>Polar distance: 12º 2/5 |
| τῶν ἐν τῇ ἑπομένῃ πλευρᾷ ὁ νότιοϲ.<br>The southern [star] of those in the trailing side[38] | β UMi | **1.4.2** οἱ λαμπρότατοι καὶ ἡγούμενοι τῶν ἐν τῷ πλινθίῳ ταύτηϲ ἀϲτέρων<br>the brightest and leading ones of the stars in its quadrangle (β UMi and γ UMi)<br><br>**3.5.10** οἱ λαμπρότατοι καὶ ἡγούμενοι τῶν ἐν τῷ πλινθίῳ τεϲϲάρων<br>the brightest and leading [stars] of the four [stars] in the quadrangle (β UMi, γ UMi) | RA: **1.7.11**: the end of the Scorpion<br><br>RA (**3.5.10**)<br>beginning of the Archer (240º = 16h00m) | Stars in the shoulder (β UMi and γ UMi)<br><br>The northern one of the two [stars] in the shoulders<br><br>RA = beginning of Archer = RA 240º<br><br>polar distance: 40º [rather: 8º] |
| τῆϲ αὐτῆϲ πλευρᾶϲ ὁ βόρειοϲ<br>The northern [star] of the same side | γ UMi | **1.4.2** οἱ γὰρ λαμπρότατοι καὶ ἡγούμενοι τῶν ἐν τῷ πλινθίῳ ταύτηϲ ἀϲτέρων,<br>the brightest and leading ones of the stars in its quadrangle β UMi, γ UMi);<br><br>**3.5.10** οἱ λαμπρότατοι καὶ ἡγούμενοι τῶν ἐν τῷ πλινθίῳ τεϲϲάρων<br>the brightest and leading [stars] of the four [stars] in the quadrangle (β UMi, γ UMi); | RA (**3.5.10**)<br>Beginning of the Archer (240º = 16h00m) | Stars in the shoulder (β UMi and γ UMi)<br><br>RA = beginning of Archer = RA 240º |

There is a quite good match, even if sometimes there are small discrepancies—which however are also present in the *Exegesis* itself, where β UMi is given two slightly different right ascensions: once (1.7.11) it is said to be at the end of the Scorpion and once is said to at the beginning of the Archer (3.5.10)—which is however more of a different way of expressing the same right ascension since the end of the Scorpion almost coincides with the beginning of the Archer.

As already noticed, if the description in AL goes back to Hipparchus, as seems very likely, then for Hipparchus β UMi was not the head and γ UMi was not the front foot of the Small Bear, as instead they were for Eudoxus and Aratus. Rather, Hipparchus seems to have placed the shoulders on these two stars. This of course begs the question of where the head and the front feet were according to him: perhaps they were not marked by any star, but just imagined. In fact, for β UMi and γ UMi to be the shoulders, Hipparchus had to imagine the Bear as either seen from above (a very odd position, yet he had a similar view for the Dragon[39]) or completely turned toward us (more likely, since this is also Ptolemy's depiction—see figure 3b). In fact, Hipparchus might have wanted to change the 'anatomical shape' of the Small Bear compared to the one of Eudoxus and Aratus because, if the head was indeed β UMi and the front feet were γ UMi, the figure would have been even more out of proportion when measured against the very long tail ending with α UMi. Thus, he might have imagined a bigger body to try to correct the overall shape. Even so (and in Ptolemy's asterism as well) the tail is still too long for a bear—however the Hipparchan Small Bear is still an improvement compared to the mini-bear with an extra-long tail of Eudoxus and

[38] Ptolemy calls it the 'trailing' side since he uses ecliptic coordinates; for Hipparchus who uses equatorial coordinates, this is instead the 'leading' side.

[39] See Hipp. 1.4.4-5 and Schironi (forthcoming)-a, ad loc.

Aratus.[40] In fact, both Hipparchus and Ptolemy seem to prefer to describe the ‘body’ of the Small Bear just as the quadrangle (β UMi, γ UMi, η UMi, ζ UMi). However, if this fragment from AL goes back to Hipparchus, he might have wanted to use more anatomically descriptive labels, perhaps in line with the ‘old tradition’ (in Aratus and Eudoxus the two Bears were, as noted before, just bears, not quadrangles), improving the anatomical shape—something that also Ptolemy claims to be doing in *Synt.* 7 before introducing his own catalogue of stars.[41]

After having analyzed the fragment in AL, we can now look at PAberd. 12. This is the entry as edited by Savio (with her apparatus, in a reduced form:

[μο]ί̣ρα̣ϲ. ̣[?      [
[πρὸϲ τὸν] βόρι̣ο[ν ἡ] ἄρ̣[κτοϲ μικρὰ ἔχει ἀϲτέραϲ ζ̅·]
ὁ βόριο]ϲ̣ δύο τῶ̣ν ἡγουμ̣[ένων ἀϲτέρων λαμπ-]
[ρῶν τῶ]ν̣ ἐ̣ν πλινθείῳ̣, ὅϲ[ ἐϲτιν ἐπὶ κεφαλῆϲ, ἀπέ-]
[χει τοῦ] πό̣λου μο̣ίραϲ η̅ λι̣π̣[ούϲαϲ μικρῷ̣· ὁ νοτι-]
[ώτατ]ο̣ϲ δὲ̣ ὁ ἄκρα τῇ οὐρ̣ᾳ̣[ ἀϲτήρ, ὅϲ ἐϲτιν]
[ἔϲχατο]ϲ̣, ἀπέχει τοῦ β̣ορίου[ πόλου μοίραϲ ιβ̅]
[καὶ δύο π]εντημόρ[ι]α.      [      ]

2 l. βόρειον
3 l. βόρειοϲ
4 l. πλινθίῳ
5 l. λειπούϲαϲ?
6 l.ὁ ⟨ἐν⟩ ἄκρᾳ τῇ οὐρ̣ᾷ
7 l. βορείου
8 l. πεμπτημόρια

Degrees?
[toward the] northern (pole) the Sm[all Bear has 7 stars].
[The north]ern one of the two lead[ing bright]
[stars] in the quadrangle, whi[ch is on the head, (β UMi) is]
[al]most 8 degrees away from the pole; [the most]
[southern] one, the [star] on the tip of the tail[ (α UMi), which is
[the last] one, is [12 degrees and 2/]5 away from the pole.

---

[40] If this is correct and Hipparchus considered β UMi and γ UMi to be the shoulders, then he was talking of the old image of the Bear when he said: Hipp. 1.5.7 φανερὸν δὲ μᾶλλον γίνεται τὸ λεγόμενον ὑφ’ ἡμῶν ἐπὶ τῆϲ Μικρᾶϲ Ἄρκτου. <u>ὁμολογουμένωϲ</u> γὰρ ἐπ’ ἐκείνηϲ ἥ τε κεφαλὴ καὶ οἱ πόδεϲ ἐν τοῖϲ τέϲϲαρϲιν ὑπ’ αὐτῶν τίθενται τοῖϲ τὸ πλινθίον περιέχουϲιν· [Our argument becomes more evident in the case of the Small Bear. For, f<u>ollowing the general consensus,</u> in it the head and the feet are placed by them in the four [stars] (β UMi, γ UMi, η UMi, ζ UMi) which outline the quadrangle].

[41] Synt. 1.2, 37.11 Heiberg καὶ ταῖϲ διαμορφώϲεϲι δ' αὐταῖϲ ταῖϲ καθ' ἕκαϲτον τῶν ἀϲτέρων οὐ πάντωϲ ϲυγκεχρήμεθα ταῖϲ αὐταῖϲ, αἷϲ καὶ οἱ πρὸ ἡμῶν, καθάπερ οὐδ' ἐκεῖνοι ταῖϲ ἔτι πρὸ αὐτῶν, ἀλλ' ἑτέραιϲ πολλαχῇ κατὰ τὸ οἰκειότερον καὶ μᾶλλον ἀκόλουθον τῷ εὐρύθμῳ τῶν διατυπώϲεων, οἷον ὅταν, οὓϲ ὁ Ἵππαρχοϲ ἐπὶ τῶν ὤμων τῆϲ Παρθένου τίθηϲιν, ἡμεῖϲ ἐπὶ τῶν πλευρῶν αὐτῆϲ κατονομάζωμεν διὰ τὸ μεῖζον αὐτῶν φαίνεϲθαι τὸ πρὸϲ τοὺϲ ἐν τῇ κεφαλῇ διάϲτημα τοῦ πρὸϲ τοὺϲ ἐν τοῖϲ ἀκροχείροιϲ, τὸ δὲ τοιοῦτον ταῖϲ μὲν πλευραῖϲ ἐφαρμόζειν, τῶν δὲ ὤμων παντάπαϲιν ἀλλότριον εἶναι. [Furthermore, for the individual [stars] of the constellations we have not used the descriptions which our predecessors [used], just as they did not use those of their own predecessors; often, [we have used] different [ descriptions] according what is more natural and more in line with the proportions of those figures, as for example when we call the stars which Hipparchus places on the shoulders of the Maiden, those on their sides, because their distance toward the [stars] in the head appears to be greater than their [distance] toward the [stars] in the tip of the hand, and this fits the sides while it is totally inconsistent with the shoulders].

Several data have been supplemented by Savio based on other sources, especially AL. The supplement for the polar distance of the star on the tip of the tail (α UMi) is quite certain, due to part of the fraction being preserved (καὶ δύο π]εντημόρ[ι]α). Moreover, the definition of α UMi as ὁ ἄκρα τῇ οὐρᾷ̣[ ἀςτήρ] matches Hipparchus' definition.[42]

In the papyrus we can read only one other number with certainty: the 8 degrees of polar distance of one of the two leading stars of the quadrangle. This value corresponds with the polar distance of β UMi at Hipparchus' time, as seen above. It also confirms the suggested emendation for the text in AL above. Hipparchus also described β UMi as one of the two leading stars in the quadrangle (the other being γ UMi) (Hipp. 1.4.2 and 3.5.10) just like in the papyrus. However, this is not how AL describes β UMi, since in AL β UMi has only an anatomical description ("The northern one of the two [stars] in the shoulders"). In fact, an anatomical description for β UMi is also present in the papyrus, where, if we accept Savio's reading, we read:

> ὁ βόριο]ς̣ δύο τῶ̣ν ἡγουμ̣[ένων ἀςτέρων λαμπ-]
> [ρῶν τῶ]ν̣ ἐ̣ν πλινθείῳ̣, ὅ̣ς[ ἐςτιν ἐπὶ κεφαλῆς, ἀπέ-]
> [χει τοῦ] πό̣λου μο̣ίρας η̅ λ̣ι̣π̣[ούςας μικρῷ
>
> [The north]ern one of the two lead[ing bright]
> [stars] in the quadrangle, whi[ch is on the head, is]
> [al]most 8 degrees away from the pole;

In Savio's supplement, β UMi is positioned 'on the head', which corresponds to the asterism of Aratus and Eudoxus, according to Hipparchus. In fact, in 1.7.11 Hipparchus also calls β UMi as the 'head' of the Small Bear.[43] It might be that also in this entry in the papyrus Hipparchus was following the 'general consensus', which is mentioned in Hipp. 1.5.7 (footnote 40). However, if the supplement in PAberd 12 is correct, then the AL text must be corrupt since there β UMi is defined as the star in the shoulders. But one could also supply the papyrus otherwise:

> ὁ βόριο]ς̣ δύο τῶ̣ν ἡγουμ̣[ένων ἀςτέρων λαμπ-]
> [ρῶν τῶ]ν̣ ἐ̣ν πλινθείῳ̣, ὅ̣ς[οι ἐν τοῖς ὤμοις, ἀπέ-]
> [χει τοῦ] πό̣λου μο̣ίρας η̅ λ̣ι̣π̣[ούςας μικρῷ
>
> [The north]ern one of the two lead[ing bright]
> [stars] in the quadrangle, [stars] whi[ch are in the shoulders, is]
> [al]most 8 degrees away from the pole;

this would correspond exactly to what we read in AL:

> item *aquilonium* [pro: aquilonius?] duobus, quibus in humero distat aquilonio polo sortes *XL* [pro: VIII?].
>
> The northern one of the two [stars] in the shoulder is *40* [rather: 8?] degrees away from the northern pole.

---

[42] Hipp. 1.5.19 τοῦ ἐν ἄκρᾳ τῇ οὐρᾷ τῆς Ἄρκτου [in the tip of the tail of the [Small] Bear]

[43] Hipp. 1.7.11: ἡ μὲν γὰρ κεφαλὴ τῆς Μικρᾶς Ἄρκτου ἐπέχει κατὰ τὸν παράλληλον τῷ ἰςημερινῷ κύκλον τὴν τελευτὴν τοῦ Cκορπίου· [For the head of the Small Bear (β UMi) occupies the end of the Scorpion along the circle parallel to the equator].

Accordingly, while *PAber.* 12 has both a geometrical and an anatomical description for β UMi (which were both known to Hipparchus), AL eliminated the geometrical one and kept only the anatomical one.

## 4 The Aratus Latinus, PAberd 12, and the Dragon

The Dragon too is described both by AL and by PAberd 12. AL has the following text:[44]

> Serpens, qui ad aquilonium iacet, in longitudine quidem habet incidens circulum aequinoctialis circuli a quarta sorte Leonis *quod* [pro: usque?] XXVIII Capricornii.
> in latitudine autem habet sortes XXVI et *duas partes* [rather: tertiam partem?]. In unam quidem partem poli *qui* [pro: quae?] ad Leonem sortes *V* [pro: X?] et duas partes, in alio vero quod ad Sagittarium sortes *XXXV* [pro: XXXVII?].
>
> praecedit quidem et qui ad summitatem caudae habens a Leone *sortem unam* [pro: sortes tres?].
> Extremus ad orientem iacet aquilonius, de quibus a latere. […]
> qui super collo ab *alterutrum* [pro: alterutro?[45]] scriptus distat a polo sortes *XXVII* [pro: XXXVII?].
> itemque *aput* [pro: apud?] Leonis quidem partes *australis* [pro: borealis] in cauda distans ab hoc polo sortes X.
>
> The Serpent, which lies toward the north, in length occupies an arc of the equinoctial circle, cutting it, from the fourth degree of the Lion *which* [rather: to] 28 [degrees] of Capricorn.
> In width it occupies 26 degrees and *two thirds* [rather: one third?]. In the part of the sky which is in the Lion [it occupies] *5* [rather 10?] degrees and two thirds [from the pole], in the other [section] which is in the Archer [it occupies] *35* [rather: 37?] degrees [from the pole].
>
> The [star] which is also in the tip of the tail leads (i.e. is the most western one), occupying *1 degree* [rather: 3 degrees?] from the Lion.
> The last [star] to the east lies in the north, among those(?) in the side. […]
> The [star] which is marked on the neck by one of the two(?)[46] is *27* [rather: 37?] degrees away from the pole.
> And the southern [rather: northern] [star] in the tail, in the parts of the Lion, is 10 degrees away from this pole.

Here too we have discrepancies between the limits given in the first part and those marked by the stars.

<u>Boundaries</u>

---

[44] For the original edition (with no indications of possible lacunae or errors), see Maass 1898, 189; for a summary of the content, see Neugebauer 1975, 290.

[45] as in the revised Aratus Latinus.

[46] "The [star] which is marked on the neck by one of the two". The meaning of this phrase is hard to grasp; the passage is clearly corrupted but I have no convincing emendation to suggest. I wonder whether it might refer to the fact that γ Dra (the star meant here—see below) is the southern limit but also a star that 'grazes' the horizon at the latitude of Greece, since there the Dragon is a circumpolar constellation. Thus, when γ Dra touches the horizon, it is both rising and setting; a similar concept is expressed by Aratus: Phaen. 61-62: Κείνη που κεφαλὴ τῇ νίϲϲεται, ἧχί περ ἄκραι / μίϲγονται δύϲιέϲ τε καὶ ἀντολαὶ ἀλλήλῃϲιν [That head [i.e., the Dragon's head] traverses the region where the limits of the settings and the risings are mixed with each other]. The other option is that it refers to the fact that the neck of the Dragon is close to the foot of the Kneeler or Hercules (e.g., 'The [star] on the neck marked by Hercules' foot")—yet I have no suggestion about how and from which original reading one could generate 'ab alterutrum', the reading attested in the manuscripts.

- W-E extension (right ascension): [175 degrees], from the fourth degree of the Lion (123º) to Capricorn 28º (298º)
- N-S extension (polar distance): 26 degrees and 2/3 [rather: 1/3?], from 5º [rather: 10º?] and 2/3 from the pole to 35º [rather: 37º] [from the pole]

Stars Marking the Boundaries

- W: the [star] in the tip of the tail: Lion 1º [rather: 3º?] (right ascension) = RA 121º [rather: RA 123º]
- E: The last [star] to the east, in the north, among those in the side; [ … right ascension missing]
- N: The southern [rather: northern] [star] in the tail, in the region of the Lion: 10º (polar distance) [= decl. +80º]
- S: The [star] on the neck: 27º [rather: 37º?] (polar distance) [= decl. + 63º [rather: +53º]]

The total for the N-S extension as preserved in the manuscripts is clearly incorrect and does not match the data in the second section. My suggestion to correct this passage depends once again on the stars marking those boundaries. We can start from the E-W limit.

According to Hipparchus, the star on the tip of the tail of the Dragon is λ Dra.[47] In 129 BCE it had a RA of 123º 02′. In fact, Hipparchus in the *Exegesis* gives its RA as "ca. 3º of the Lion" which corresponds to 123º.[48] This value also corresponds to what is expressed in the first part of AL with ordinal numbers as 'the fourth degree of the Lion'; thus the value attached in the second part (Lion 1º) is probably corrupted and should be changed into 3º Lion (an omission with number correction, from "sortes III" to "sortem I", or just a misreading of the letter numeral in Greek: from γ̄ μοι TO ᾱ μοι?).

The 'northern star' on the side is more difficult to track; among the stars belonging to the Dragon, the closest is ε Dra; Hipparchus does not mention it in the *Exegesis*, so we cannot compare his description to the one in AL. However, in 129 BCE ε Dra had a RA of 295º 39′, which matches rather well the value provided by AL in the constellation's general limit: Capricorn 28º (298º), with a mistake of ca. 2.5º. It does also respect the description of the 'northern star on the side' (figure 4). Indeed, λ Dra and ε Dra are the two longitudinal extremes of the Dragon.

As for the southern and northern limits, the star on the tail marking the northern limit can still be λ Dra; in 129 BCE it had a declination of 79º 32′, which indeed matches the value of AL for this star (10º from the pole = decl. +80º). Of course, this star cannot be described as the 'southern' star in the tail; it must be a mistake for 'northern'. Gysembergh at al.[49] suggest κ Dra, which had a declination of 81º 27′ in 129 BCE. While this is indeed the most northern one, both stars (both mentioned by Hipparchus in the *Exegesis*) could work in terms of polar distance. However, the right ascension of κ Dra in 129 BCE was 154º 21′, which in Hipparchus' "zodiacal" terminology would be Maiden 5º. However, AL clearly says that the star in question is in the region of the Lion, which proves that the star is λ Dra. Moreover, in the *Exegesis* Hipparchus reports Eudoxus claiming that the 'topmost star' of the Dragon was on top of the head of the Bear:

> **1.2.3** Ἐν δὴ τούτῳ τῷ συντάγματι Εὔδοξος περὶ μὲν τοῦ Δράκοντος οὕτως γράφει (fr. F 15)· "μεταξὺ δὲ τῶν Ἄρκτων ἐστὶν ἡ τοῦ Ὄφεως οὐρά, τὸν ἄκρον ἀστέρα ὑπὲρ τῆς κεφαλῆς ἔχουσα τῆς Μεγάλης Ἄρκτου. …"

---

47 Hipp. 2.6.9 and 2.6.10; 3.2.11 ὁ ἐν ἄκρᾳ τῇ οὐρᾷ [the [star] in the tip of the tail].

48 Hipp. 1.5.4 ὁ μὲν γὰρ ἐν ἄκρᾳ τῇ οὐρᾷ τοῦ Δράκοντος ἐπέχει ὡς κατὰ παράλληλον κύκλον {τοῦ} Λέοντος μοίρας γ̄ [… For the [star] within the tip of the tail of the Dragon (λ Dra) occupies ca. 3 degrees of the Lion along the circle parallel [to the equator]].

49 Gysembergh, Williams and Zingg 2022, 389.

In this treatise, Eudoxus writes about the Dragon as follows (fr. F 15): “Between the Bears there is the tail of the Serpent [i.e., the Dragon], which has its topmost star over the head of the Great Bear. …”.

Since according to Hipparchus, Eudoxus (and Aratus) identified the head of the Great Bear with α UMa,[50] the star above it is λ Dra, and not κ Dra. To conclude, both κ Dra and λ Dra would work in terms of declination, but both the right ascension of κ Dra and the testimony of Hipparchus’ *Exegesis* prove that the star is question is actually λ Dra.

As for the southern limit, in the *Exegesis* Hipparchus says that the most southern star of the Dragon is γ Dra which is however identified as the southern or left temple.[51] Unlike Ptolemy, who depicts the head of the Dragon in profile, both Aratus and Hipparchus considered it as facing us (Aratus) or the outside of the sphere (Hipparchus).[52] If one sees the Dragon’s head in this way, indeed the temple can coincide with the neck—allowing in this case for a different description of this star compared to Hipparchus’ description in the *Exegesis*. Indeed, γ Dra is also the southern limit of the Dragon. As for polar distance or declination, in 129 BCE γ Dra had a declination of 52º 56′; in fact, in the *Exegesis* (1.4.8 and 1.113) Hipparchus gives it as a polar distance of 37º, which corresponds to a declination of +53º. AL gives a polar distance of 27º, when listing the stars, and 35º, when measuring the boundaries. There is clearly an error here. If the original was 37º, then 27 could have arisen from a paleographical confusion in the Greek ($\overline{\kappa\zeta}$ = 27 instead of the correct $\overline{\lambda\zeta}$ = 37) or the Latin (XXVII instead XXXVII). Also the 35º could have originated from an original 37º in Greek ($\overline{\lambda\varepsilon}$ instead of $\overline{\lambda\zeta}$) or in Latin (from XXXVII to XXXV); in both cases the mistake seems easier to occur in Latin. Restoring this value seems correct not only because it is the same polar distance provided by Hipparchus in the *Exegesis* but also because the total extension N-S as provided by AL would be more correct, since the original has 26º and 2/3—where the 2/3 is clearly a mistake from 1/3 in subtracting 10º and 2/3 from 37º.

The identifications are thus the following:[53]

Stars Marking the Boundaries: Identifications and Data for 129 BCE

- W: the [star] in the tip of the tail (λ Dra): Lion 1º [rather: 3º] (right ascension) = RA 121º [rather: RA 123º= first part in AL] = *Exegesis* ≈ RA 129 BCE: 123º 02′
- E: The last [star] to the east, in the north, among those in the side (ε Dra); [ … right ascension missing] [right ascension in first part: Capricorn 28º = RA 298º] ≈ RA 129: 295º 39′,
- N: The southern [rather: northern] [star] in the tail (λ Dra): 10º (polar distance) [= decl. +80º] ≈ RA 129: 79º 32′

---

[50] Hipp. 1.5.2 ὅτι δὲ ἀγνοοῦϲιν, ἐκ τούτων ἐϲτὶ φανερόν. ἡ μὲν γὰρ κεφαλὴ τῆϲ Μεγάληϲ Ἄρκτου κατὰ τοὺϲ προειρημένουϲ ἄνδραϲ ἐϲτὶν ὁ βορειότεροϲ ἀϲτὴρ τῶν δύο τῶν ἡγουμένων ἐν τῷ πλινθίῳ, ἐπὶ δὲ τῶν ἐμπροϲθίων ποδῶν κεῖται ὁ νοτιώτεροϲ τῶν αὐτῶν ἀϲτέρων. 1.5.3 ὅτι μὲν γὰρ ὁ προειρημένοϲ ἀϲτὴρ κατ’ αὐτοὺϲ ἐπὶ τῆϲ κεφαλῆϲ κεῖται, φανερὸν τοῦτο ποιοῦϲιν ἐκ {τε} τοῦ λέγειν τὸν ἐν ἄκρᾳ τῇ οὐρᾷ τοῦ Δράκοντοϲ ἀϲτέρα κεῖϲθαι κατὰ τὴν κεφαλὴν τῆϲ Ἄρκτου. [But it is clear that they are mistaken from the following points. For the head of the Great Bear, according to the aforementioned authors, is the more northern star (α UMa) of the two leading ones (α UMa, β UMa) in the quadrangle, while the more southern (β UMa) of these stars lies on the front feet. They make it clear that for them the aforementioned star (α UMa) lies on the head by saying that the star within the tip of the tail of the Dragon (λ Dra) lies near the head of the Bear].

[51] Hipp 3.2.1ὁ νοτιώτεροϲ κρόταφοϲ [the more southern temple]; 1.11.3: ὁ γὰρ νοτιώτεροϲ αὐτῶν καὶ ἐπὶ τοῦ ἀριϲτεροῦ κροτάφου κείμενοϲ [the one that is more southern and that lies on the left temple].

[52] See discussion in Schironi (forthcoming)-a, ad Hipp. 1.4.5-6.

[53] The same identifications were suggested by Neugebauer 1975, 291 and Gysembergh, Williams and Zingg 2022, 389—with the exception, for the latter, of κ Dra for the northern limit.

- S: The [star] on the neck (γ Dra): 27º [rather: 37º] (polar distance) [= decl. + 63º [rather: +53º]] = *Exegesis* (polar distance) ≈ RA 129: 52º 56′

To conclude, a synopsis of the stars mentioned here with the data present in Ptolemy, Hipparchus and AL will make this clearer:

| Ptolemy | Star | Hipparchus in Exegesis | Coordinates given in Exegesis | AL |
|---|---|---|---|---|
| ὁ ἐπάνω τῆς κεφαλῆς<br>The [star] above the head | γ Dra | **1.4.4** ὁ ἀριστερὸς κρόταφος<br>The left temple<br><br>**1.4.8:** ὁ δὲ νότιος κρόταφος<br>the southern temple<br><br>**3.2.1:** ὁ νοτιώτερος κρόταφος<br>the more southern temple<br><br>**1.11.3** ὁ νοτιώτερος αὐτῶν καὶ ἐπὶ τοῦ ἀριστεροῦ κροτάφου κείμενος<br>the more southern of them [i.e. the Dragon’s stars] and the one that lies on the left temple | Polar distance (**1.4.8, 1.11.3**): 37º | The [star] on the neck<br><br>Polar distance: 27º/35º [rather: 37º] |
| τῆς ἑπομένης πλευρᾶς ὁ βόρειος<br>Of the trailing side, the northern [star] | ε Dra | | | The last [star] to the east lies to the north, among those in the side<br><br>RA: Capricorn 28º = 298º |
| ὁ λοιπὸς καὶ ἐπ' ἄκρας τῆς οὐρᾶς<br>The remaining star and on the tip of the tail | λ Dra | **1.5.3, 1.5.4, 2.6.10, [2.6.16], 3.2.11** ὁ ἐν ἄκρᾳ τῇ οὐρᾷ<br>The [star] within the tip of the tail | RA: **(1.5.4):** ca. 3º Lion | the [star] in the tip of the tail<br><br>the southern [rather: northern] [star] in the tail<br><br>RA: 4th degree of the Lion = 3º Lion<br><br>[AL] Polar distance: 10º from NP [= decl. +80º] |

PAberdeen 12 has just the beginning of the entry on the Dragon:

[μετὰ δὲ] ὁ διὰ τῶν ἄρκτων ὄ̣[φις ἔχει τὴν οὐρὰν
[διὰ τῶ]ν̣ ἄρκτων φερο[μένην,νεύουσαν δὲ τὴν κε-]
[φαλὴν ἕ]ω̣ς ἔνπ̣ρ̣ο̣ς̣θ̣[ε τοῦ ἐνγουνασι? …

11 l. ἔμπροσθε

[Afterward] the S[erpent] between the Bears [has the tail]
[which is] carried [between] the Bears[, while the head inclines]
[to]ward the front [of the Kneeler …?]

In the fragment, there are no extension limits nor coordinates, thus it cannot be checked against the values correct for Hipparchus’ time. The description according to which this Serpent lies between the Bears is typical of the early descriptions of this constellation, for example in Aratus,[54] and does not necessarily point to Hipparchus. Yet the naming of the constellation as ὁ διὰ τῶν ἄρκτων ὄ̣[φιc], 'the S[erpent] between the Bears' is indeed attested in Hipparchus (Hipp. 1.11.1, 1.11.3, 3.3.7) as well as in Eudoxus (Hipp. 1.2.11).

The description of PAberd. 12 thus does not necessarily suggest that it is Hipparchus' since there are no specific data that can clearly point to his time, but it does offer a description of the Dragon in line with Hipparchus. On the other hand, the description in AL does match Hipparchus' data in the *Exegesis*.

## 5 Hipparchus' Catalogue (CCR, AL, PAberd. 12): Content, Transmission, and Abridgments

The evidence offered by PAberd. 12, AL, and, above all, CCR indeed point to a description of constellations that ultimately goes back to Hipparchus. Not only do many of the coordinates match those of the time of Hipparchus and/or given by Hipparchus himself in the *Exegesis.* In addition, also the way right ascension is expressed (with an 'ecliptic' or 'zodiacal' nomenclature) in these fragments is attested only in Hipparchus. Similarly, polar distance is the equatorial latitude preferred by Hipparchus also in the *Exegesis* (even if there he also gives declinations[55]). Moreover, the description and names of these constellations (e.g. the "Serpent between the Bears" or the stars of the Crown counted on the basis of α CrB) are generally those attested also in Hipparchus' *Exegesis*.

CCR preserves the most detailed description, with the 'most scientific' presentation of the data. Even if there are some discrepancies when compared to similar content in Hipparchus' *Exegesis*, the fragment in CCR is probably the closest to the original. Compared with CCR, both AL and PAberd 12 are poor copies of it. In the structure, AL is quite close to CCR's format. Yet the gross linguist mistakes as well as the many problems with the numerical data points to a less-than-precise tradition. Indeed, this is part of the Aratus Latinus, that is, the eight-century collection of scholia and other exegetical material to Aratus, which is clearly a very damaged and impoverished version of the original Aratean material.

PAberd. 12, on the other hand, is just a scrap of a papyrus; yet if we compare the two entries on the Small Bear, an additional problem emerges in the version preserved by the papyrus:

> Minor Arcturus, qui ad aquilonem adiacet, in longitudine quidem habet quartam partem aequinoctialis circuli et sortes *septem* [pro: septemdecim?].
> a latere autem habet sortes *una* [pro: quattuor?] *semis* [pro: duas quintas partes?], quae ab octo sortibus minus sunt duodecim *pars* [pro: sortes?] et *duo quinta sors* [pro: duae quintae partes?].

---

[54] Phaen. 45-52: Τὰc δὲ δι' ἀμφοτέρας οἵη ποταμοῖο ἀπορρὼξ / εἰλεῖται, μέγα θαῦμα, Δράκων, περί τ' ἀμφί τ' ἐαγὼc / μυρίος· αἱ δ' ἄρα οἱ σπείρης ἑκάτερθε φέρονται / Ἄρκτοι, κυανέου πεφυλαγμέναι ὠκεανοῖο. / Αὐτὰρ ὅγ' ἄλλης μὲν νεάτῃ ἐπιτείνεται οὐρῇ, / ἄλλην δὲ σπείρῃ περιτέμνεται· ἡ μέν οἱ ἄκρη / οὐρὴ πὰρ κεφαλὴν Ἑλίκης ἀποπαύεται Ἄρκτου· / σπείρῃ δ' ἐν Κυνόcουρα κάρη ἔχει· [Through both of them [i.e., the Bears] the Dragon winds, like a river's flow, a great marvel, with an endless broken course around both sides; the Bears are borne on both sides of its coil, avoiding the dark Ocean. [The Dragon] stretches over one [of the Bears] with the tip of its tail and cuts off the other with its coil. The tip of its tail stops at the head of the Bear Helice [i.e., the Great Bear], and Cynosura [i.e. the Small Bear] has its head in its coil].

[55] See Schironi (forthcoming)-a, Chapter 5.

antecedunt autem illam quae deducunt *XII* [pro: II?], de quibus in humerum habent principatum de Sagittario.
novissima ad orientem adiacet ad summum habens Piscem sortes XVII.
<u>item *aquilonium* [pro: aquilonius?] duobus, quibus in humero distat aquilonio polo sortes *XL* [pro: VIII?].</u>
restant aequinoctiales.
**australis quoque est, qui in cauda.**
reliqui vero duo nominantur Circenses, eo quod in circuitu perambulent.

The Little Bear, which lies toward the north, in length has the fourth part of the equinoctial circle and *7* [rather: 17?] degrees.
And from the side (i.e. in width) it has *1* [rather: 4?] degrees and *½* [rather: 2/5?], which are 12 degrees and 2/5 less of 8 degrees.

There are *12* [rather 2?] [stars] which lead in it, of which those in the shoulder occupy the beginning of the Archer.
The very last one to the east lies to the extreme [limit], occupying the Fish, 17 degrees.
<u>The northern one of the two [stars] in the shoulder is *40* [rather: 8?] degrees away from the northern pole.</u>
There remain […] equinoctial […].
**The southern [star] is also the one in the tail.**
The two remaining [stars] are called ‘Circenses’ because they go around a circle.

[μο]ί̣ρα̣ϲ. ̣[?        [
[πρὸϲ τὸν] βόρ̣ι̣ο[ν ἡ] ἄρ̣[κτοϲ μικρὰ ἔχει ἀϲτέραϲ ζ̅·]
<u>ὁ βόριο]ϲ̣ δύο τῶ̣ν ἡγουμ̣[ένων ἀϲτέρων λαμπ-]</u>
<u>[ρῶν τῶ]ν̣ ἐ̣ν πλινθείῳ̣, ὅϲ[οι ἐν τοῖϲ ὤμοιϲ, ἀπέ-]</u>[56]
<u>[χει τοῦ] πό̣λου μο̣ίραϲ η̅ λι̣π̣[ούϲαϲ μικρῷ̣·</u> **ὁ νοτι-]**
**[ώτατ]ο̣ϲ δὲ̣ ὁ ἄκρα τῇ̣ οὐρ̣ᾷ̣[ ἀϲτήρ, ὅϲ ἐϲτιν]**
**[ἔϲχατο]ϲ̣, ἀπέχει τοῦ β̣ορίου[ πόλου μοίραϲ ι̅β̅]**
**[καὶ δύο π]εντημόρ[ι]α.**        [            ]

[De]grees?
[toward the] northern (pole) the Sm[all Bear has 7 stars].
<u>[The north]ern one of the two lead[ing bright]</u>
<u>[stars] in the quadrangle, whi[ch are in the shoulders, is]</u>
<u>[al]most 8 degrees away from the pole;</u> **[the most]**
**[southern] one, the [star] on the tip of the tail[, which is**
**[the last] one, is [12 degrees and 2/]5 away from the pole.**

I have underlined the data concerning the star marking the northern limit; in bold are the data concerning the star marking the southern limit. As is clear, the text as preserved in the papyrus covers only the last part of the entry in AL. Of course, this is just a scrap, and we could speculate that the first part of the entry was preserved in the missing part. However, if we follow Savio's reconstruction of the first two lines, this is a very reduced version, actually even different from the one preserved in CCR and AL. According to Savio's text, there is a brief 'presentation' of the Small Bear with the total number of stars (l. 2: [toward the] northern (pole) the Sm[all Bear has 7 stars]), a detail which is absent in CCR and in the AL fragments. After providing the total number of stars in the Small Bear, the papyrus only provides the northern and southern limits, with the two stars marking them. Not only are the stars marking the eastern and western limits not mentioned, but also the first part of this entry, in which the E-W and N-S total extensions in degrees are given, is missing. In other words, the papyrus could contain an even more abridged version of the original. Yet I wonder whether one can supplement the first two lines in a different way, so that, for example, they turn out to be dealing with the star marking the eastern boundary—something which would correspond to what we read in AL:

> novissima ad orientem adiacet ad summum habens Piscem sortes XVII.

> The very last one to the east lies to the extreme [limit], occupying the Fish, 17 degrees.

[56] Here in l. 4 I am adopting my own suggested supplement, ὅϲ[οι <u>ἐν τοῖϲ ὤμοιϲ rather than Savio's</u> ὅϲ[ ἐϲτιν <u>ἐπὶ κεφαλῆϲ</u>.

In Hipparchus' Greek this would correspond to something such as:

> Ἔϲχατοϲ δὲ πρὸϲ ἀνατολὰϲ κεῖται, ἐπέχων Ἰχθύων μοίραϲ ι̅ζ̅.

Indeed, in line 1 there are traces compatible with [μο]ί̣ρα̣ϲ. In the second line it is tempting to read a word connected with 'north' (βορέαϲ, βορρᾶϲ, βόρειοϲ) and given that the omicron before the first gap is clear, the most likely solution is one form of the adjective βόρειοϲ, 'northern', as suggested by Savio. If there is a reference to 'north' however, line 2 is most likely *not* discussing the eastern limit. Still, instead of giving a presentation of the entire constellation as having 7 stars as Savio reconstructs it, this line could introduce the passage to the stars marking the N-S limit. A possible integration could be:

> [πρὸϲ τὸν] βόρι̣ο[ν ἡ] ἄρ̣[κτοϲ μικρὰ κεῖται.]

> The Small Bear lies toward the north.

The line would be shorter (31 letters) than the one proposed by Savio (37 letters); however, also line 6 has 31 letters, and the others have 33 (l. 3), 36 (l. 4 according to her supplement; 34 letters according to mine), 36 (l. 5) and 35 (l. 7) letters, so 31 letters in l. 2 is in fact not so impossible. One could object that saying that the constellation lies toward the north is less informative than providing the number of stars in a catalogue—especially if star coordinates are given in the following lines. This is certainly true but has the advantage of allowing another part of the entry still dealing with the Small Bear to precede line 2, for example the part discussing the E-W boundary, with the statement that the constellation was located toward the north in l. 2, a fitting introduction to the discussion of the N-S limits. If instead we go with Savio's proposal (which, I agree, is a good one) we must also accept that this papyrus would be an oddity compared to the other fragments of this Star Catalogue, because, unlike the fragments in CCR and AL, it provided the total star numbers of each constellation (which nor CCR or AL do) and also did not even list all the limits but only the N-S boundaries. In fact, even if indeed l. 2 is the beginning of the entry, my proposed restoration would better fit the evidence from the other fragments in CCR and AL, which always start by saying in which portion of the celestial sphere the constellation lies. In this case, simply PAberd 12 would have omitted (because of a lacuna in the original?) the total E-W and N-S extension as well as the stars marking the E-W limits. Yet it would have preserved the same incipit as the AL and CCR and would not have another set of data (the total number of stars) which neither CCR or AL have.

No matter what we make of the first two lines, the papyrus offers an interesting testimony: unlike CCR, which is clearly set out as a scientific text for insiders, and AL, which, though very corrupted, was still part of an exegetical corpus to an astronomical 'classic', namely, Aratus'

*Phaenomena*, PAberd 12 is a papyrus dating to the 2nd or 3rd century CE and coming from the Fayum. In other words, PAberd. 12 testify to the use of these texts beyond the restricted circle of professional astronomers or commentators of Aratus. Of course, the papyrus could be part of a larger text commenting on Aratus—yet as far as our evidence goes, it rather suggests a text containing only descriptions of constellations one after the other.

However, if we accept Savio's reconstruction, it is a very poor copy of its model, since it misses key data to allow mapping constellations out. Such a text was unlikely to be a text of professional astronomers. Rather, it must have been a personal copy of someone who was interested in constellations and their technical side but did not care much about having all the data.

The circulation of 'less than precise' scholarly or scientific texts in the Egyptian *chōra* is indeed attested in other genres. For example, among grammatical and scholarly papyri (that is, papyri containing commentaries on or editions of Greek literary authors) this phenomenon is very much attested. The 'mark' of scholarly editions and commentaries going back to the Alexandrian grammarians is the presence of critical signs. These were specific signs added to the left margin of the 'scholarly' edition and also repeated in the commentary, which was contained in a separate roll—in this way the critical signs could work as a reference and link between the 'critical editions' (*ekdoseis*) and the scholarly, self-standing commentary (*hypomnēma*).[57] While we have examples of these scholarly texts, probably owned by teachers or professional philologists, we also have 'deluxe copies' of the *Iliad* or *Odyssey*, which were not books owned by scholars but a precious possession to show off the owner's literacy. In some of these copies we have *some* critical signs added to *some* lines. This is the case of the Hawara Homer, a luxury copy of Book 2 of the *Iliad*, dating back to the second century CE. In this beautifully written text sparse marginal notes and critical signs are added, here and there. When compared to similar and more scholarly papyri covering the same portion of text (e.g. *POxy* 1086) and manuscripts (the *Venetus A*), we see that the Hawara Homer shows the critical signs in the correct position (that is, next to the specific lines they refer to); yet they are far less than in those other scholarly copies. Similarly, its (short) notes are just a summary of the longer notes preserved in *POxy* 1086 or in the scholia in the *Venetus* A.[58] This could be the type of situation represented by PAberd 12 compared to AL and, even more so, CCR. From this fragment it looks like the copy in the papyrus had only stars marking the north and southern limits of constellations but not those marking the E-W boundaries. It was thus a 'shorter' version of more technical texts, which people with some intellectual aspirations might have liked to possess. Yet a 'star boundary catalogue' with only the most northern and most southern stars would be almost useless for any serious practitioner.

However, given the uncertainties of the reconstruction of the first two lines, we should also consider the possibility that what remains in PAberd 12 is just the end of the entry similar to those preserved by CCR and AL, and that the W-E and S-N boundaries as well as the stars marking the W and E limits have been lost.

## 6 Hipparchus' Catalogue (CCR, AL, PAberd 12) and Hipparchus' Exegesis

The preceding analysis has shown that the fragments preserved by CCR, AL, and PAberd. 12 contain Hipparchan elements, not only because the coordinates they preserve align with those of

---

[57] See Schironi 2018, 49-62.

[58] See McNamee 1981, 253; Schironi 2021, 261-264.

Hipparchus' time; more importantly, these fragments have some overlaps both in terms of data and style with the *Exegesis*, which is the only work by Hipparchus transmitted by direct tradition. However, the stellar astronomy of the *Exegesis* presents also some additional elements which are different from the stellar catalogue as emerges from the fragments in CCR, AL, and PAberd. 12.

The analysis in the *Exegesis*, and especially the Catalogue of Simultaneous Risings and Setting (2.4.1-3.4.12) and the Appendix on the 1-hour meridians (3.5.1-23), both presuppose a Catalogue of Stars with (equatorial) coordinates. In the *Exegesis* Hipparchus provides equatorial coordinates of specific stars; he also describes specific stars by their position in the shape of the constellation itself. Yet there Hipparchus describes and gives coordinates for stars that do not mark the 'limits' of constellations, as the Catalogue recovered primarily from CCR suggests. Just to cover the constellations discussed in this article, polar distances are provided for β UMa (1.11.5), γ UMa (1.11.5), μ Dra? (uncertain identification) (1.4.8), β Dra (1.4.8); right ascensions are provided for β UMa (1.5.12), γ UMa (1.5.9 and 1.5.11), δ UMa (1.5.11), θ UMa (3.5.1b), δ UMa (3.5.4). In other words, in the *Exegesis* there are coordinates which do not belong only to the stars marking constellations' limits. In order to have those data, Hipparchus must have had another catalogue collecting all the stars for each constellation, each with equatorial coordinates.

Such a catalogue is also suggested by the comparison with Ptolemy's Catalogue (Books 7 and 8 of the *Syntaxis*). Indeed, the description of some stars by Hipparchus is very close to the description of the same stars in Ptolemy—sometimes verbatim. If indeed Ptolemy adapted Hipparchus catalogue for his own Catalogue,[59] we need to suppose that there was an Hipparchan Catalogue organized as the one of Ptolemy.

# 7 Conclusions

The analysis of the fragment of Hipparchus' Star Catalogue preserved in CCR, AL and PAberd 12 vis-à-vis the stellar data present in Hipparchus' *Exegesis*, suggests the following conclusions.

CCR, AL, and PAberd12 do preserve fragments of a star catalogue, whose date coincide with the epoch of Hipparchus; also, the style for expressing coordinates ("zodiacal" right ascension and polar distance), the star descriptions and some coordinates do match Hipparchus' practice and data in the *Exegesis*. These fragments thus most likely derive from a Catalogue of Stars by Hipparchus, a catalogue focused on constellations' boundaries. However, in the *Exegesis* Hipparchus preserves more coordinates of stars that do not mark constellation limits. This suggests the existence of at least another catalogue with coordinates for *all* the stars belonging to a constellation.

We could speculate that the fragment in CCR is just the beginning of the Catalogue's entry on the Northern Crown, and that after the description of the constellation's boundaries, there was the description of each star in it, with the equatorial coordinates (lost instead in the excerpts of AL, as well as in PAberd. 12, which clearly ends with the N-S limit of the Small Bear). Or we could imagine that Hipparchus put together different star catalogues, with different purposes. The one preserved in CCR, AL and PAberd. 12 was a catalogue focused on the boundaries of constellations (a 'Constellation-Boundaries Catalogue'). Then Hipparchus might have wanted to compose a more full-fledged catalogue with a list of every star belonging to each constellation and with equatorial coordinates for all stars. This (a 'Full Catalogue') is the catalogue that Ptolemy used, turning the equatorial coordinates into ecliptic coordinates.[60] In particular, the 'Full Catalogue' (with equatorial coordinates) seems to have been already available to Hipparchus when he composed the *Exegesis*.

[59] The discussion about Ptolemy's use (or appropriation) of Hipparchus' Catalogue of stars for his own Catalogue is notoriously a fraught topic. A clear summary of the debate is in Evans 1998, 264-274; see also Neugebauer 1975, 280-288 and Pedersen and Jones 2011, 253-258. A more detailed reconstruction of all the studies on this question is in Graßhoff 1990, 23-91.

While this latter hypothesis might seem less economical than supposing only one catalogue with first the constellation boundaries and then the list of stars with descriptions and coordinates, it has in fact two advantages: first, all the witnesses of the catalogue with boundaries do not show any remain of a second part with star descriptions. Second, we have evidence that Hipparchus toyed with different types of star catalogues: in the second part of the *Exegesis* there is a long 'Catalogue of Simultaneous Risings and Settings', followed by a 'Catalogue of Stars Marking the 24 1-Hour Meridians'. These two are also catalogues of stars, even though in a form and with data that are different from the other two catalogues discussed here, the 'Constellation-Boundaries Catalogue' and the 'Full Catalogue' used by Ptolemy.

Indeed because of the importance of stars and constellations in antiquity, it is not far-fetched to think that there were different types of star catalogues, each serving its own purpose.[61] For Hipparchus, four catalogues are thus attested—and by 'catalogues' I mean lists of stars organized according to specific principles and with different formats, but not necessarily lists including all known stars:

1. The 'Constellation-Boundaries Catalogue' (preserved by CCR, AL and PAberd 12)
2. The 'Full Catalogue' with all the stars, their descriptions and equatorial coordinates (which was used for compiling the *Exegesis* and is also at the basis of Ptolemy's Catalogue)
3. The 'Catalogue of Simultaneous Risings and Settings' (in *Exegesis* 2.4.1-3.4.12)
4. the 'Catalogue of Stars Marking The 24, 1-hr Meridians' (in *Exegesis*, 3.5.1-23).

While the last three Catalogues have a specific content and scope, one might wonder the reason for a 'Constellation-Boundaries Catalogue' (no. 1). We could imagine that this was just a first step toward a more full-fledged star catalogue (no. 2), with a less complete descriptions of constellations. Hipparchus knew already that stars in constellations could be listed and described one by one, within the constellation's shape, since Eratosthenes' *Catasterismi* had already adopted this format in the third century BCE. But he might have wanted to prepare a more precise one with all the coordinates—and the 'Constellation-Boundaries Catalogue' might have been a first step toward the more complete 'Full Catalogue'.

A 'Constellation-Boundaries Catalogue' however might have been useful for other purposes. For example, it could have been used for a first, rough making of globes, at least non-professional ones. Using this catalogue when preparing a star map or a globe, globe-makers could have found out the boundaries of each constellation and used them as guidelines to draw their figures. This idea seems to be supported by the evidence of ancient globes. Of the three globes preserved from antiquity (the Farnese globe, [62] the Kugel/Paris globe[63] and the Mainz globe,[64] dating between 150 BCE and 200 CE) only the Mainz globe has stars in it, and they seem also to be placed almost randomly. But the Farnese globe and the Kugel globe do not have any stars, which suggests that single stars did not seem to have been very important in ancient depictions of constellations, at least for globes not used by professional astronomers. As long as one knew the general shape of the constellation and the area occupied by it, this was enough for having a rough idea of a constellation. Globe-makers were then free to draw the constellation shapes they pleased, provided that they were contained in the constellations' own boundaries, which could be checked in the 'Constellation-Boundaries Catalogue' This might also explain why we have such a different variety of iconographies for constellations in our globes.

[60] Unless we supposed that Hipparchus made a third catalogue when he discovered precession, all in ecliptic coordinates.

[61] Another type of catalogue organized in a different way and for different purposes (stars are reference for planetary motions) is the Catalogue of Stars in the Handy Tables of Ptolemy (T 20A).

Another option, which seems to be suggested by PAberd 12 is that the 'Constellation-Boundaries Catalogue' might have been more popular than a full-fledged Catalogue of the type of Ptolemy's Catalogue—in fact we have no papyrus fragment preserving the latter, which is explicable by the nature of such a text, clearly only for professionals, hence without a wide readership among laypeople. On the other hand, the 'Constellation-Boundaries Catalogue' must have been much more compressed and shorter than a complete star catalogue (Ptolemy's Catalogue occupies two books of the *Syntaxis*!). Yet, even if in a reduced format, such catalogue could have had the pretentions of 'serious astronomy' because, unlike the catalogue of Eratosthenes' *Catasterismi*, it had celestial coordinates. In addition, by providing the 'geometrical' shapes (triangles or quadrangles) of a constellation by stating the stars marking those boundaries (in the three cases preserved, the Great Bear is contained in a quadrangle, the Small Bear and the Dragon are contained in a triangle) it might have given the impression that one could indeed use it to locate constellations in the sky. In fact, there is some evidence in papyri that the use of geometrical shapes to map-out the different constellations was indeed in use among laypeople. PParis 1, vii 11- viii 4, the famous *Ars Eudoxi* (P.Par. 1 – P.Louvre N 2325 + N 2388),[65] indeed introduces constellations in this way, outlining triangles, quadrangles, and circles. In some cases, these stars mark two or more constellations relative to each other, for example Orion, the Dog and Procyon:

> P.Paris 1 vii 12-19:
> τρίγωνα δὲ $^{v}$ καὶ τετράγωνα $^{v}$ καὶ
> κύκλους $^{v}$ καὶ εὐcι̣είας $^{v}$ γραμμὰc τὰ ἄcτρα
> cυντηροῦcιν $^{v}$ ἐν τῆι πε̣[ρ]ι̣φορᾶι τῆc c⟨φ⟩αίραc.
> ὅ τε γὰρ τοῦ ὠρίωνοc $^{v}$ ὦμοc $^{v}$ καὶ ὁ κύων
> καὶ ὁ προκύων $^{v}$ τρίγωνόν $^{v}$ τι νοεῖται.
>
> The constellations preserve their triangles, squares, circles and straight lines in the revolution of the (celestial) sphere. For example, the shoulder of Orion and the Dog and the Lead-Dog (i.e. Procyon) are conceived of as a kind of triangle.

In other cases, more similar to our case, the stars mark one specific constellation, and they work like (rough) boundaries; for example:

vii 19-21

---

[62] On this globe, see Thiele 1898, 19-42; Valerio 1987; Valerio 2005; Schaefer 2005; Duke 2006; Dekker 2013, 84-102, 111-115

[63] On this globe, see Cuvigny 2004; Dekker 2013, 57-69, 102-106

[64] On this globe, see Künzl 1997/1998; Künzl 2000; Dekker 2013, 69-80, 106-111.

[65] See Jones and Schironi forthcoming.

τετράγωνα δὲ
οἱ ἀπὸ τοῦ $^v$ ἵππου $^v$ ἀςτέρες $^v$ καὶ ἀπὸ
τῶν ἄρ[κ]των. [

But the stars of the Horse
and those of the Bears (display) a square.

viii 2-4

κύκλ̣ο̣υς δὲ οἱ ἀπὸ τοῦ $^v$ ςτεφάνου
ἀςτέ̣ρ̣ες $^v$ καὶ ἀπὸ τῆς $^v$ κεφαλῆς
τοῦ λέοντος.

The stars from the Crown and
(those) from the head
of the Lion are circles.

However, there is an essential difference between singling out some stars that trace a geometrical shape within a constellation (as, for example, the quadrangle of the Small Bear, β UMi, γ UMi, η UMi, ζ UMi) or within three or more constellations as in the case of the Dog (α CMa), Procyon (α CMi) and Orion (α Ori). These are usually bright stars (especially in the latter case) and are relatively close to each other, with no other bright stars in the middle, which should cloud the geometrical shape. This is not the case for stars marking constellation boundaries as in Hipparchus' Catalogue here analyzed: first, one cannot choose them, since these stars are selected not on the basis of their brightness but on the basis of their marking limits, so some might be not so bright; second, if the constellation is large (e.g. Heracles or Andromeda), they might be quite far apart from each other, so that it would be quite hard for anyone to recognize the geometrical shapes of the boundaries when sky gazing. Thus, I doubt that the 'Constellation-Boundaries Catalogue' by Hipparchus would have been useful for recognizing 'constellation areas' in the sky by laypeople. Yet, such a catalogue with *some* coordinates and stemming from such a high-level tradition, the Hipparchan one, could certainly make an impression in the library of would-be intellectuals, without taking too much space. This is why, perhaps, it was preserved in PAberd. 12.

Yet such a catalogue probably went out of fashion relatively early on, at least for the general public. This seems to be suggested by the type of evidence we have for it. Both AL and, as far as we can tell, PAberd. 12 only contain fragments for circumpolar constellations, the two Bears and the Dragon. Neugebauer, who did not know of CCR, thought that the reason was that for circumpolar constellations one could not use a description for simultaneous risings and settings.[66] This is absolutely true; in fact, these three constellations are not included in Hipparchus' Catalogue of Simultaneous Risings and Settings in the *Exegesis*; similarly, these three constellations are not mentioned in parapegmata since they never rise or set. However, the fragment in CCR suggests that the three circumpolar constellations were not the only constellations described in this way (as probably Neugebauer thought). The fact that the fragments preserved in the Aratean corpus and in papyri only pertain these three constellations might indeed be explained by assuming that at a certain point, when the 'Full Catalogue' became more widespread, the 'Constellation-Boundaries Catalogue' became superfluous, *except* for these three constellations which could also not be described in the Catalogue of Simultaneous Risings and Settings or in the more popular *parapegmata*. This is why fragments concerning these circumpolar constellations have been preserved in two different copies, the only testimonies beyond the palimpsest in CCR, and which are the more popular type of sources: a papyrus for non-professionals and scholia in the Aratean corpus connected with rather elementary teaching (while CCR could be indeed one of the last copies of the old catalogue for connoisseurs).

[66] This is what one can read in Neugebauer' notebooks: "The reason for giving this description of the 3 circumpolar constellations is probably the fact that for them definitions by simultaneous rising and setting could not be used"

Since the ancients relied so much for calendrical matters on stars, it should not surprise us that they organized stellar data in different way, for different matters. Stars were used to calculate time at night and to keep track of farming seasons; they could also be used as reference stars for tracking planetary motions, or to indicate directions. The fragments of Hipparchus indeed suggest that he tried to provide different stellar catalogues for different readers and different purposes; yet at the end, only the Full Catalogue, reworked and improved by Ptolemy, imposed itself as 'the' Catalogue of Stars. This might have occurred when stars became less crucial for daily life, so that they could only be the object of a more comprehensive and scientific catalogue, while the more partial and practical catalogues had lost most of their utility.

Francesca Schironi, University of Michigan